\pdfoutput=1
\documentclass[11pt,twoside,a4paper,pdftex,cmspaper,final,collab]{cms-tdr}

\begin{document}\cmsNoteHeader{CFT-09-017}
%
%
%

%
%
\hyphenation{env-iron-men-tal}
\hyphenation{had-ron-i-za-tion}
\hyphenation{cal-or-i-me-ter}
\hyphenation{de-vices}
%
%
\RCS$Revision: 1.10 $
\RCS$Date: 2010/01/14 16:33:18 $
\RCS$Name:  $
%
%
%

\newcommand {\etal}{\mbox{et al.}\xspace} 
\newcommand {\ie}{\mbox{i.e.}\xspace}     
\newcommand {\eg}{\mbox{e.g.}\xspace}     
\newcommand {\etc}{\mbox{etc.}\xspace}     
\newcommand {\vs}{\mbox{\sl vs.}\xspace}      
\newcommand {\mdash}{\ensuremath{\mathrm{-}}} 

\newcommand {\Lone}{Level-1\xspace} 
\newcommand {\Ltwo}{Level-2\xspace}
\newcommand {\Lthree}{Level-3\xspace}

\providecommand{\ACERMC} {\textsc{AcerMC}\xspace}
\providecommand{\ALPGEN} {{\textsc{alpgen}}\xspace}
\providecommand{\CHARYBDIS} {{\textsc{charybdis}}\xspace}
\providecommand{\CMKIN} {\textsc{cmkin}\xspace}
\providecommand{\CMSIM} {{\textsc{cmsim}}\xspace}
\providecommand{\CMSSW} {{\textsc{cmssw}}\xspace}
\providecommand{\COBRA} {{\textsc{cobra}}\xspace}
\providecommand{\COCOA} {{\textsc{cocoa}}\xspace}
\providecommand{\COMPHEP} {\textsc{CompHEP}\xspace}
\providecommand{\EVTGEN} {{\textsc{evtgen}}\xspace}
\providecommand{\FAMOS} {{\textsc{famos}}\xspace}
\providecommand{\GARCON} {\textsc{garcon}\xspace}
\providecommand{\GARFIELD} {{\textsc{garfield}}\xspace}
\providecommand{\GEANE} {{\textsc{geane}}\xspace}
\providecommand{\GEANTfour} {{\textsc{geant4}}\xspace}
\providecommand{\GEANTthree} {{\textsc{geant3}}\xspace}
\providecommand{\GEANT} {{\textsc{geant}}\xspace}
\providecommand{\HDECAY} {\textsc{hdecay}\xspace}
\providecommand{\HERWIG} {{\textsc{herwig}}\xspace}
\providecommand{\HIGLU} {{\textsc{higlu}}\xspace}
\providecommand{\HIJING} {{\textsc{hijing}}\xspace}
\providecommand{\IGUANA} {\textsc{iguana}\xspace}
\providecommand{\ISAJET} {{\textsc{isajet}}\xspace}
\providecommand{\ISAPYTHIA} {{\textsc{isapythia}}\xspace}
\providecommand{\ISASUGRA} {{\textsc{isasugra}}\xspace}
\providecommand{\ISASUSY} {{\textsc{isasusy}}\xspace}
\providecommand{\ISAWIG} {{\textsc{isawig}}\xspace}
\providecommand{\MADGRAPH} {\textsc{MadGraph}\xspace}
\providecommand{\MCATNLO} {\textsc{mc@nlo}\xspace}
\providecommand{\MCFM} {\textsc{mcfm}\xspace}
\providecommand{\MILLEPEDE} {{\textsc{millepede}}\xspace}
\providecommand{\ORCA} {{\textsc{orca}}\xspace}
\providecommand{\OSCAR} {{\textsc{oscar}}\xspace}
\providecommand{\PHOTOS} {\textsc{photos}\xspace}
\providecommand{\PROSPINO} {\textsc{prospino}\xspace}
\providecommand{\PYTHIA} {{\textsc{pythia}}\xspace}
\providecommand{\SHERPA} {{\textsc{sherpa}}\xspace}
\providecommand{\TAUOLA} {\textsc{tauola}\xspace}
\providecommand{\TOPREX} {\textsc{TopReX}\xspace}
\providecommand{\XDAQ} {{\textsc{xdaq}}\xspace}

\newcommand {\DZERO}{D\O\xspace}     


\newcommand{\de}{\ensuremath{^\circ}}
\newcommand{\ten}[1]{\ensuremath{\times \text{10}^\text{#1}}}
\newcommand{\unit}[1]{\ensuremath{\text{\,#1}}\xspace}
\newcommand{\mum}{\ensuremath{\,\mu\text{m}}\xspace}
\newcommand{\micron}{\ensuremath{\,\mu\text{m}}\xspace}
\newcommand{\cm}{\ensuremath{\,\text{cm}}\xspace}
\newcommand{\mm}{\ensuremath{\,\text{mm}}\xspace}
\newcommand{\mus}{\ensuremath{\,\mu\text{s}}\xspace}
\newcommand{\keV}{\ensuremath{\,\text{ke\hspace{-.08em}V}}\xspace}
\newcommand{\MeV}{\ensuremath{\,\text{Me\hspace{-.08em}V}}\xspace}
\newcommand{\GeV}{\ensuremath{\,\text{Ge\hspace{-.08em}V}}\xspace}
\newcommand{\TeV}{\ensuremath{\,\text{Te\hspace{-.08em}V}}\xspace}
\newcommand{\PeV}{\ensuremath{\,\text{Pe\hspace{-.08em}V}}\xspace}
\newcommand{\keVc}{\ensuremath{{\,\text{ke\hspace{-.08em}V\hspace{-0.16em}/\hspace{-0.08em}c}}}\xspace}
\newcommand{\MeVc}{\ensuremath{{\,\text{Me\hspace{-.08em}V\hspace{-0.16em}/\hspace{-0.08em}c}}}\xspace}
\newcommand{\GeVc}{\ensuremath{{\,\text{Ge\hspace{-.08em}V\hspace{-0.16em}/\hspace{-0.08em}c}}}\xspace}
\newcommand{\TeVc}{\ensuremath{{\,\text{Te\hspace{-.08em}V\hspace{-0.16em}/\hspace{-0.08em}c}}}\xspace}
\newcommand{\keVcc}{\ensuremath{{\,\text{ke\hspace{-.08em}V\hspace{-0.16em}/\hspace{-0.08em}c}^\text{2}}}\xspace}
\newcommand{\MeVcc}{\ensuremath{{\,\text{Me\hspace{-.08em}V\hspace{-0.16em}/\hspace{-0.08em}c}^\text{2}}}\xspace}
\newcommand{\GeVcc}{\ensuremath{{\,\text{Ge\hspace{-.08em}V\hspace{-0.16em}/\hspace{-0.08em}c}^\text{2}}}\xspace}
\newcommand{\TeVcc}{\ensuremath{{\,\text{Te\hspace{-.08em}V\hspace{-0.16em}/\hspace{-0.08em}c}^\text{2}}}\xspace}

\newcommand{\pbinv} {\mbox{\ensuremath{\,\text{pb}^\text{$-$1}}}\xspace}
\newcommand{\fbinv} {\mbox{\ensuremath{\,\text{fb}^\text{$-$1}}}\xspace}
\newcommand{\nbinv} {\mbox{\ensuremath{\,\text{nb}^\text{$-$1}}}\xspace}
\newcommand{\percms}{\ensuremath{\,\text{cm}^\text{$-$2}\,\text{s}^\text{$-$1}}\xspace}
\newcommand{\lumi}{\ensuremath{\mathcal{L}}\xspace}
\newcommand{\Lumi}{\ensuremath{\mathcal{L}}\xspace}
%
\newcommand{\LvLow}  {\ensuremath{\mathcal{L}=\text{10}^\text{32}\,\text{cm}^\text{$-$2}\,\text{s}^\text{$-$1}}\xspace}
\newcommand{\LLow}   {\ensuremath{\mathcal{L}=\text{10}^\text{33}\,\text{cm}^\text{$-$2}\,\text{s}^\text{$-$1}}\xspace}
\newcommand{\lowlumi}{\ensuremath{\mathcal{L}=\text{2}\times \text{10}^\text{33}\,\text{cm}^\text{$-$2}\,\text{s}^\text{$-$1}}\xspace}
\newcommand{\LMed}   {\ensuremath{\mathcal{L}=\text{2}\times \text{10}^\text{33}\,\text{cm}^\text{$-$2}\,\text{s}^\text{$-$1}}\xspace}
\newcommand{\LHigh}  {\ensuremath{\mathcal{L}=\text{10}^\text{34}\,\text{cm}^\text{$-$2}\,\text{s}^\text{$-$1}}\xspace}
\newcommand{\hilumi} {\ensuremath{\mathcal{L}=\text{10}^\text{34}\,\text{cm}^\text{$-$2}\,\text{s}^\text{$-$1}}\xspace}


\newcommand{\zp}{\ensuremath{\mathrm{Z}^\prime}\xspace}


\newcommand{\kt}{\ensuremath{k_{\mathrm{T}}}\xspace}
\newcommand{\BC}{\ensuremath{{B_{\mathrm{c}}}}\xspace}
\newcommand{\bbarc}{\ensuremath{{\overline{b}c}}\xspace}
\newcommand{\bbbar}{\ensuremath{{b\overline{b}}}\xspace}
\newcommand{\ccbar}{\ensuremath{{c\overline{c}}}\xspace}
\newcommand{\JPsi}{\ensuremath{{J}/\psi}\xspace}
\newcommand{\bspsiphi}{\ensuremath{B_s \to \JPsi\, \phi}\xspace}
\newcommand{\AFB}{\ensuremath{A_\mathrm{FB}}\xspace}
\newcommand{\EE}{\ensuremath{e^+e^-}\xspace}
\newcommand{\MM}{\ensuremath{\mu^+\mu^-}\xspace}
\newcommand{\TT}{\ensuremath{\tau^+\tau^-}\xspace}
\newcommand{\wangle}{\ensuremath{\sin^{2}\theta_{\mathrm{eff}}^\mathrm{lept}(M^2_\mathrm{Z})}\xspace}
\newcommand{\ttbar}{\ensuremath{{t\overline{t}}}\xspace}
\newcommand{\stat}{\ensuremath{\,\text{(stat.)}}\xspace}
\newcommand{\syst}{\ensuremath{\,\text{(syst.)}}\xspace}

\newcommand{\HGG}{\ensuremath{\mathrm{H}\to\gamma\gamma}}
\newcommand{\gev}{\GeV}
\newcommand{\GAMJET}{\ensuremath{\gamma + \mathrm{jet}}}
\newcommand{\PPTOJETS}{\ensuremath{\mathrm{pp}\to\mathrm{jets}}}
\newcommand{\PPTOGG}{\ensuremath{\mathrm{pp}\to\gamma\gamma}}
\newcommand{\PPTOGAMJET}{\ensuremath{\mathrm{pp}\to\gamma +
\mathrm{jet}
}}
\newcommand{\MH}{\ensuremath{\mathrm{M_{\mathrm{H}}}}}
\newcommand{\RNINE}{\ensuremath{\mathrm{R}_\mathrm{9}}}
\newcommand{\DR}{\ensuremath{\Delta\mathrm{R}}}


\newcommand{\PT}{\ensuremath{p_{\mathrm{T}}}\xspace}
\newcommand{\pt}{\ensuremath{p_{\mathrm{T}}}\xspace}
\newcommand{\ET}{\ensuremath{E_{\mathrm{T}}}\xspace}
\newcommand{\HT}{\ensuremath{H_{\mathrm{T}}}\xspace}
\newcommand{\et}{\ensuremath{E_{\mathrm{T}}}\xspace}
\newcommand{\Em}{\ensuremath{E\!\!\!/}\xspace}
\newcommand{\Pm}{\ensuremath{p\!\!\!/}\xspace}
\newcommand{\PTm}{\ensuremath{{p\!\!\!/}_{\mathrm{T}}}\xspace}
\newcommand{\ETm}{\ensuremath{E_{\mathrm{T}}^{\mathrm{miss}}}\xspace}
\newcommand{\MET}{\ensuremath{E_{\mathrm{T}}^{\mathrm{miss}}}\xspace}
\newcommand{\ETmiss}{\ensuremath{E_{\mathrm{T}}^{\mathrm{miss}}}\xspace}
\newcommand{\VEtmiss}{\ensuremath{{\vec E}_{\mathrm{T}}^{\mathrm{miss}}}\xspace}

%

\newcommand{\ga}{\ensuremath{\gtrsim}}
\newcommand{\la}{\ensuremath{\lesssim}}
\newcommand{\swsq}{\ensuremath{\sin^2\theta_W}\xspace}
\newcommand{\cwsq}{\ensuremath{\cos^2\theta_W}\xspace}
\newcommand{\tanb}{\ensuremath{\tan\beta}\xspace}
\newcommand{\tanbsq}{\ensuremath{\tan^{2}\beta}\xspace}
\newcommand{\sidb}{\ensuremath{\sin 2\beta}\xspace}
\newcommand{\alpS}{\ensuremath{\alpha_S}\xspace}
\newcommand{\alpt}{\ensuremath{\tilde{\alpha}}\xspace}

\newcommand{\QL}{\ensuremath{Q_L}\xspace}
\newcommand{\sQ}{\ensuremath{\tilde{Q}}\xspace}
\newcommand{\sQL}{\ensuremath{\tilde{Q}_L}\xspace}
\newcommand{\ULC}{\ensuremath{U_L^C}\xspace}
\newcommand{\sUC}{\ensuremath{\tilde{U}^C}\xspace}
\newcommand{\sULC}{\ensuremath{\tilde{U}_L^C}\xspace}
\newcommand{\DLC}{\ensuremath{D_L^C}\xspace}
\newcommand{\sDC}{\ensuremath{\tilde{D}^C}\xspace}
\newcommand{\sDLC}{\ensuremath{\tilde{D}_L^C}\xspace}
\newcommand{\LL}{\ensuremath{L_L}\xspace}
\newcommand{\sL}{\ensuremath{\tilde{L}}\xspace}
\newcommand{\sLL}{\ensuremath{\tilde{L}_L}\xspace}
\newcommand{\ELC}{\ensuremath{E_L^C}\xspace}
\newcommand{\sEC}{\ensuremath{\tilde{E}^C}\xspace}
\newcommand{\sELC}{\ensuremath{\tilde{E}_L^C}\xspace}
\newcommand{\sEL}{\ensuremath{\tilde{E}_L}\xspace}
\newcommand{\sER}{\ensuremath{\tilde{E}_R}\xspace}
\newcommand{\sFer}{\ensuremath{\tilde{f}}\xspace}
\newcommand{\sQua}{\ensuremath{\tilde{q}}\xspace}
\newcommand{\sUp}{\ensuremath{\tilde{u}}\xspace}
\newcommand{\suL}{\ensuremath{\tilde{u}_L}\xspace}
\newcommand{\suR}{\ensuremath{\tilde{u}_R}\xspace}
\newcommand{\sDw}{\ensuremath{\tilde{d}}\xspace}
\newcommand{\sdL}{\ensuremath{\tilde{d}_L}\xspace}
\newcommand{\sdR}{\ensuremath{\tilde{d}_R}\xspace}
\newcommand{\sTop}{\ensuremath{\tilde{t}}\xspace}
\newcommand{\stL}{\ensuremath{\tilde{t}_L}\xspace}
\newcommand{\stR}{\ensuremath{\tilde{t}_R}\xspace}
\newcommand{\stone}{\ensuremath{\tilde{t}_1}\xspace}
\newcommand{\sttwo}{\ensuremath{\tilde{t}_2}\xspace}
\newcommand{\sBot}{\ensuremath{\tilde{b}}\xspace}
\newcommand{\sbL}{\ensuremath{\tilde{b}_L}\xspace}
\newcommand{\sbR}{\ensuremath{\tilde{b}_R}\xspace}
\newcommand{\sbone}{\ensuremath{\tilde{b}_1}\xspace}
\newcommand{\sbtwo}{\ensuremath{\tilde{b}_2}\xspace}
\newcommand{\sLep}{\ensuremath{\tilde{l}}\xspace}
\newcommand{\sLepC}{\ensuremath{\tilde{l}^C}\xspace}
\newcommand{\sEl}{\ensuremath{\tilde{e}}\xspace}
\newcommand{\sElC}{\ensuremath{\tilde{e}^C}\xspace}
\newcommand{\seL}{\ensuremath{\tilde{e}_L}\xspace}
\newcommand{\seR}{\ensuremath{\tilde{e}_R}\xspace}
\newcommand{\snL}{\ensuremath{\tilde{\nu}_L}\xspace}
\newcommand{\sMu}{\ensuremath{\tilde{\mu}}\xspace}
\newcommand{\sNu}{\ensuremath{\tilde{\nu}}\xspace}
\newcommand{\sTau}{\ensuremath{\tilde{\tau}}\xspace}
\newcommand{\Glu}{\ensuremath{g}\xspace}
\newcommand{\sGlu}{\ensuremath{\tilde{g}}\xspace}
\newcommand{\Wpm}{\ensuremath{W^{\pm}}\xspace}
\newcommand{\sWpm}{\ensuremath{\tilde{W}^{\pm}}\xspace}
\newcommand{\Wz}{\ensuremath{W^{0}}\xspace}
\newcommand{\sWz}{\ensuremath{\tilde{W}^{0}}\xspace}
\newcommand{\sWino}{\ensuremath{\tilde{W}}\xspace}
\newcommand{\Bz}{\ensuremath{B^{0}}\xspace}
\newcommand{\sBz}{\ensuremath{\tilde{B}^{0}}\xspace}
\newcommand{\sBino}{\ensuremath{\tilde{B}}\xspace}
\newcommand{\Zz}{\ensuremath{Z^{0}}\xspace}
\newcommand{\sZino}{\ensuremath{\tilde{Z}^{0}}\xspace}
\newcommand{\sGam}{\ensuremath{\tilde{\gamma}}\xspace}
\newcommand{\chiz}{\ensuremath{\tilde{\chi}^{0}}\xspace}
\newcommand{\chip}{\ensuremath{\tilde{\chi}^{+}}\xspace}
\newcommand{\chim}{\ensuremath{\tilde{\chi}^{-}}\xspace}
\newcommand{\chipm}{\ensuremath{\tilde{\chi}^{\pm}}\xspace}
\newcommand{\Hone}{\ensuremath{H_{d}}\xspace}
\newcommand{\sHone}{\ensuremath{\tilde{H}_{d}}\xspace}
\newcommand{\Htwo}{\ensuremath{H_{u}}\xspace}
\newcommand{\sHtwo}{\ensuremath{\tilde{H}_{u}}\xspace}
\newcommand{\sHig}{\ensuremath{\tilde{H}}\xspace}
\newcommand{\sHa}{\ensuremath{\tilde{H}_{a}}\xspace}
\newcommand{\sHb}{\ensuremath{\tilde{H}_{b}}\xspace}
\newcommand{\sHpm}{\ensuremath{\tilde{H}^{\pm}}\xspace}
\newcommand{\hz}{\ensuremath{h^{0}}\xspace}
\newcommand{\Hz}{\ensuremath{H^{0}}\xspace}
\newcommand{\Az}{\ensuremath{A^{0}}\xspace}
\newcommand{\Hpm}{\ensuremath{H^{\pm}}\xspace}
\newcommand{\sGra}{\ensuremath{\tilde{G}}\xspace}
\newcommand{\mtil}{\ensuremath{\tilde{m}}\xspace}
\newcommand{\rpv}{\ensuremath{\rlap{\kern.2em/}R}\xspace}
\newcommand{\LLE}{\ensuremath{LL\bar{E}}\xspace}
\newcommand{\LQD}{\ensuremath{LQ\bar{D}}\xspace}
\newcommand{\UDD}{\ensuremath{\overline{UDD}}\xspace}
\newcommand{\Lam}{\ensuremath{\lambda}\xspace}
\newcommand{\Lamp}{\ensuremath{\lambda'}\xspace}
\newcommand{\Lampp}{\ensuremath{\lambda''}\xspace}
\newcommand{\spinbd}[2]{\ensuremath{\bar{#1}_{\dot{#2}}}\xspace}

\newcommand{\MD}{\ensuremath{{M_\mathrm{D}}}\xspace}
\newcommand{\Mpl}{\ensuremath{{M_\mathrm{Pl}}}\xspace}
\newcommand{\Rinv} {\ensuremath{{R}^{-1}}\xspace}

%
%
\hyphenation{en-viron-men-tal}

\cmsNoteHeader{09-017}
\title{Aligning the CMS Muon Chambers with the Muon Alignment System during an Extended Cosmic Ray Run}

\author[cern]{The CMS Collaboration}

\date{\today}

\abstract{
The alignment system for the muon spectrometer of the CMS detector comprises three independent subsystems
of optical and analog position sensors. It aligns muon chambers with respect to each other and to the central
silicon tracker. System commissioning at full magnetic field began
in 2008 during an extended cosmic ray run.
The system succeeded in tracking muon detector movements of up to 18\,mm and rotations of several
milliradians under magnetic forces. Depending on coordinate and subsystem, the system achieved
chamber alignment precisions of 140\,-\,350\,$\mu$m and 30\,-\,200\,$\mu$rad, close to the precision
requirements of the experiment.
Systematic errors on absolute positions are estimated to be 340\,-\,590\,$\mu$m based on comparisons
with independent photogrammetry measurements.
}

\hypersetup{%
pdfauthor={CMS collaboration},%
pdftitle={Aligning the CMS Muon Chambers with the Muon Alignment System during an Extended Cosmic Ray Run},%
pdfsubject={CMS},%
pdfkeywords={CMS, muon, alignment}}

\maketitle 


\section{Introduction}

The primary goal of the Compact Muon Solenoid (CMS) experiment~\cite{CMS_detector_paper} 
is to explore particle physics at the TeV energy scale exploiting the proton-proton collisions 
delivered by the Large Hadron Collider (LHC)~\cite{LHC}. 
Precise measurement of muons up to the TeV momentum range 
requires the muon chambers to be aligned with respect to each other, and to the central tracking system, 
with an accuracy of a few hundred microns in position and about 40 microradians in orientation. 
The CMS Collaboration conducted a month-long data-taking exercise 
known as the Cosmic Run At Four Tesla (CRAFT) during October and November of 2008, 
with the goal of  commissioning the experiment for extended operation~\cite{CRAFTGeneral}. 
With all installed detector systems participating, CMS recorded 270 million 
cosmic-ray muon triggered events with the solenoid at its nominal axial field 
strength of 3.8\,T. 
This paper focuses on the muon alignment system and the results 
obtained with the data taken during the CRAFT exercise.
A separate paper describes in detail the track-based muon alignment 
procedures and results~\cite{CRAFT08_track_alignment_paper}.
A complete alignment of all muon chambers was not available for CRAFT;
the partial results shown here reflect the level of development of the system by the end of November 2008. 

\begin{figure}[hbtp]
\begin{center}
\includegraphics[width=0.8\textwidth]{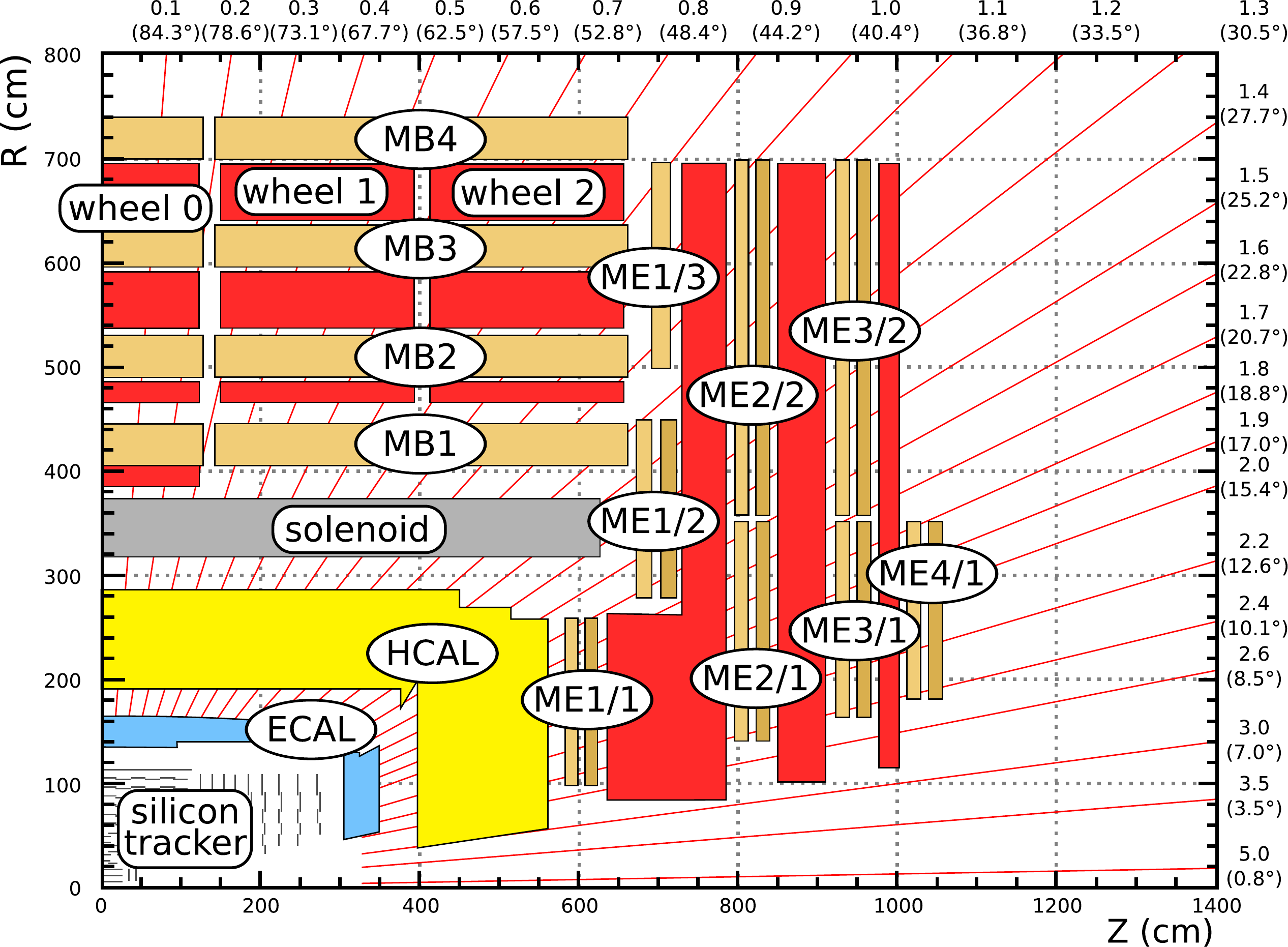}
\caption{Schematic view of a CMS quadrant.}
\label{fig:cms_quadrant}
\end{center}
\end{figure}

The central feature of the CMS apparatus is a superconducting
solenoid of 6\,m internal diameter, 13\,m length, and 4\,T magnetic field. The magnetic flux generated by
the solenoid is returned via the surrounding 1.5\,m thick, steel return yoke, approximately 22\,m
long and 14\,m in diameter, arranged as a 12-sided cylinder (each side is called a sector)
closed at each end by endcaps.
The barrel yoke is subdivided into five wheels (YB0, YB$\pm$1, and YB$\pm$2) and each
endcap yoke into three disks (YE$\pm$1, YE$\pm$2, and YE$\pm$3). 
Three technologies are used for the detection of muons: drift tubes (DT) in the central region,
cathode strip chambers (CSC) in the endcaps, and resistive plate chambers (RPC) throughout barrel and endcap.
The DT system comprises 250 chambers mounted onto the five wheels of the barrel yoke and
arranged into four concentric layers, called muon barrel (MB) stations, interleaved with the steel yoke plates. 
The CSC system consists of 468 chambers mounted on the endcap disks, perpendicular to the beam pipe, and arranged 
in four layers, called muon endcap (ME) stations, in each endcap. The RPCs are not
aligned by the muon alignment system. Figure~\ref{fig:cms_quadrant} shows the barrel wheels, endcap disks, and muon
stations for one quadrant of the CMS detector.

With the exception of the central wheel, which is fixed, the other wheels and disks are movable 
along the beam direction to allow opening the yoke for the installation and maintenance of the detectors.
Gravitational distortions lead to static deformations of the yoke elements that generate displacements of 
the muon chambers with respect to their design position of up to several millimetres. 
These displacements can be measured 
within a few hundred microns by photogrammetry when the detector is open. Similarly, 
CSC chambers can be located by photogrammetry to a few hundred microns within the disks in the plane 
perpendicular to the beam line.
However, the repositioning of the large elements of the yoke after opening and closing of the CMS detector, 
though rather precise, cannot be better than a few millimetres given their size and weight.
In addition, the magnetic flux induces huge forces that cause deformations and movements that may be as large 
as several millimetres, and must be carefully tracked by the alignment system. 
The eleven yoke elements are compressed and slightly tilted. The endcap disks are bent, 
and the central part of the YE$\pm$1 disks is deflected inward by roughly 15\,mm.
Thermal equilibrium of the yoke is reached after several months of operation, 
with thermal effects expected in the sub-millimetre range. Thermal effects will be studied
in the future when CMS runs over long periods of time under stable detector conditions.
All these displacements and deformations are either partially or totally non-reproducible, 
and their typical size is an order of magnitude larger than the desired chamber position accuracy.

In order to determine the positions and orientations of the muon chambers, 
the CMS alignment strategy combines precise survey and photogrammetry information,
measurements from an optical-based muon alignment system, and the results of 
alignment procedures based on muon tracks. 
The muon alignment system discussed in this paper is fast, independent 
of beam conditions, and can provide online monitoring of relative movements precisely.
Track-based alignment needs to accumulate a large amount of data and is, therefore, intrinsically slower, 
but it provides very accurate alignment of the muon chambers with respect to the inner tracker,
which is crucial for optimal muon momentum measurement.

The CMS alignment system consists of four independent parts. Three of these parts deal with the internal 
alignment of the tracker, DT, and CSC systems. A fourth part, called the Link system, locks all four 
parts together in a common reference frame, allows simultaneous monitoring of the barrel and endcaps, 
and monitors the displacements of heavy structures during the critical closing phase and during normal 
operation.
The muon alignment system is designed to provide continuous monitoring of the muon chamber positions in the entire 
magnetic field range between 0\,T and 4\,T, and to meet the challenging constraints of large 
radiation and magnetic field tolerance, wide dynamic range, high precision, 
and tight spatial confinement. 
The system is based on a number of precise rigid structures independently supported by the tracker and 
by each yoke element. 
These structures contain optical sensors that look at the relative positions of chambers
within the same yoke element. The connection among the structures located on the various yoke elements is possible
only when CMS is closed, and is obtained through a network of laser beams, local distance sensors, 
and digital cameras.

The aim of the muon alignment system is to provide position information of the 
detector elements with a precision comparable to the intrinsic chamber resolution. 
This information is used online for triggering purposes, and offline as a correction for track reconstruction.
The internal alignment of tracker elements is described elsewhere~\cite{CRAFT08_tracker_alignment}. 


The rest of the paper is organized as follows:
Section 2 describes the system layout and alignment strategy common to all three muon 
alignment subsystems; Section 3 gives a brief description of the data acquisition and data taking experience 
during CRAFT; Sections 4 and 5 present the alignment results for barrel and endcap muon chambers, respectively.

\section{System Layout and Geometry Reconstruction}
CMS uses a coordinate system with the origin at the
nominal collision point, the $x$-axis pointing to the centre of the
LHC, the $y$-axis pointing up (perpendicular to the LHC plane), and
the $z$-axis along the anticlockwise-beam direction. These global 
CMS coordinates are denoted as $x_{CMS}$, $y_{CMS}$, and $z_{CMS}$. 
The polar angle $\theta$ is measured from the positive $z_{CMS}$-axis, 
the azimuthal angle $\phi$ is measured in the $x_{CMS}$--$y_{CMS}$ plane, and $r_{CMS}$ is
the perpendicular distance from the beam line.
Each muon chamber and layer has its own local coordinate system,
centred on the subdetector with $z_\textrm{local}$ being perpendicular to the
measurement planes. The $x_\textrm{local}$ axis is always the direction measured with 
the highest precision, arranged to coincide with the global $r\phi$ direction at the chamber
origin, and $y_\textrm{local}$ completes a right-handed local Cartesian coordinate system.  
In the absence of misalignments, $y_\textrm{local}$ is parallel to the beam line for DT chambers 
and radial for CSC chambers. A sketch of the local coordinate systems 
can be found in Ref.~\cite{CRAFT08_track_alignment_paper}.
Since the magnetic field is parallel to $z_{CMS}$, the muon 
transverse momentum is measured from the curvature of tracks in the $r_{CMS}$--$\phi_{CMS}$ plane. 
The precision on displacements in the $r\phi$ direction and rotations of chambers around their local 
axis parallel to $z_{CMS}$ is therefore directly related to the momentum resolution. Alignment in other
degrees of freedom affects the momentum measurement as higher-order corrections.

\begin{figure}[hbtp]
\begin{center}
\includegraphics[width=0.45\textwidth]{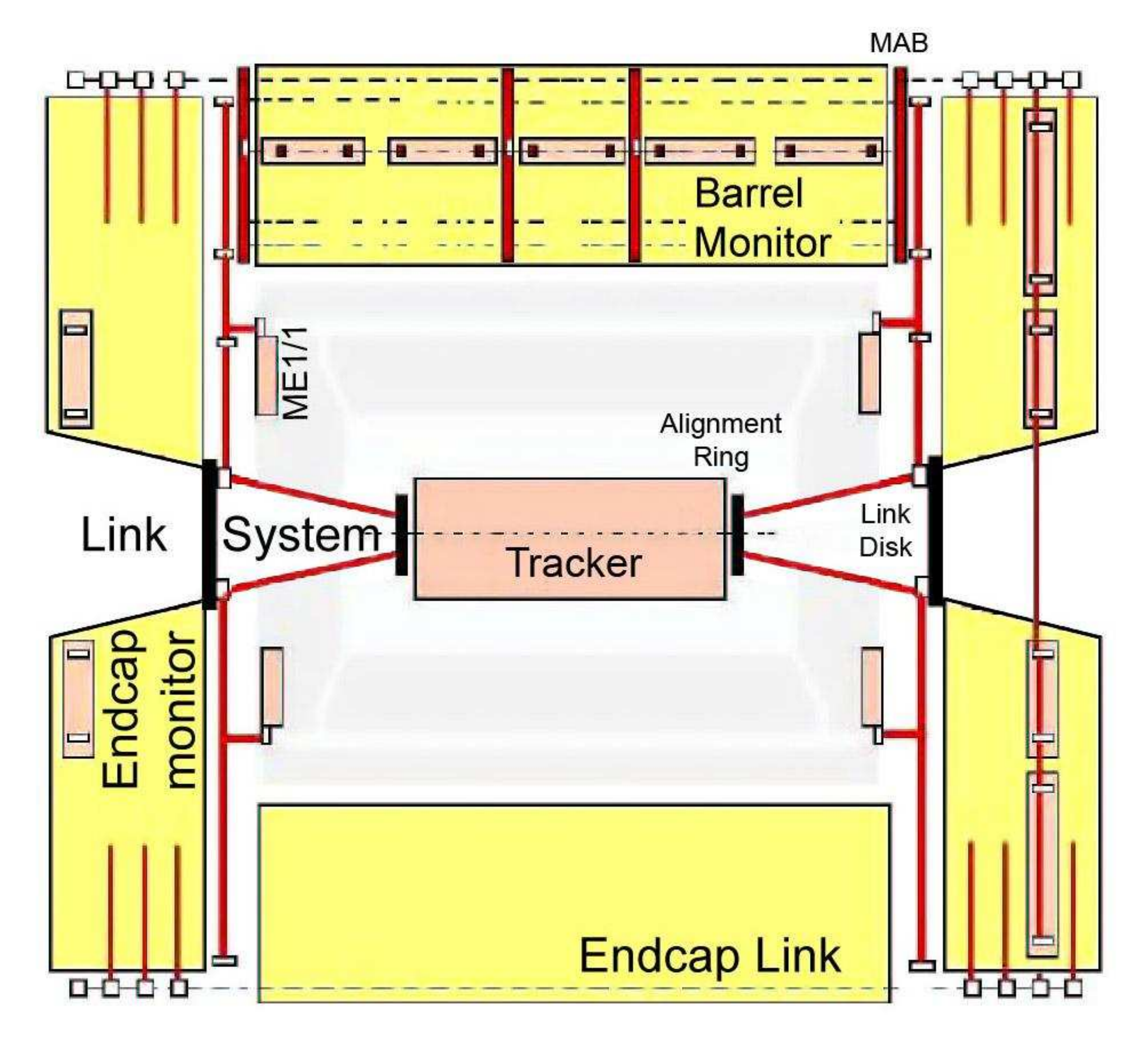}
\includegraphics[width=0.45\textwidth]{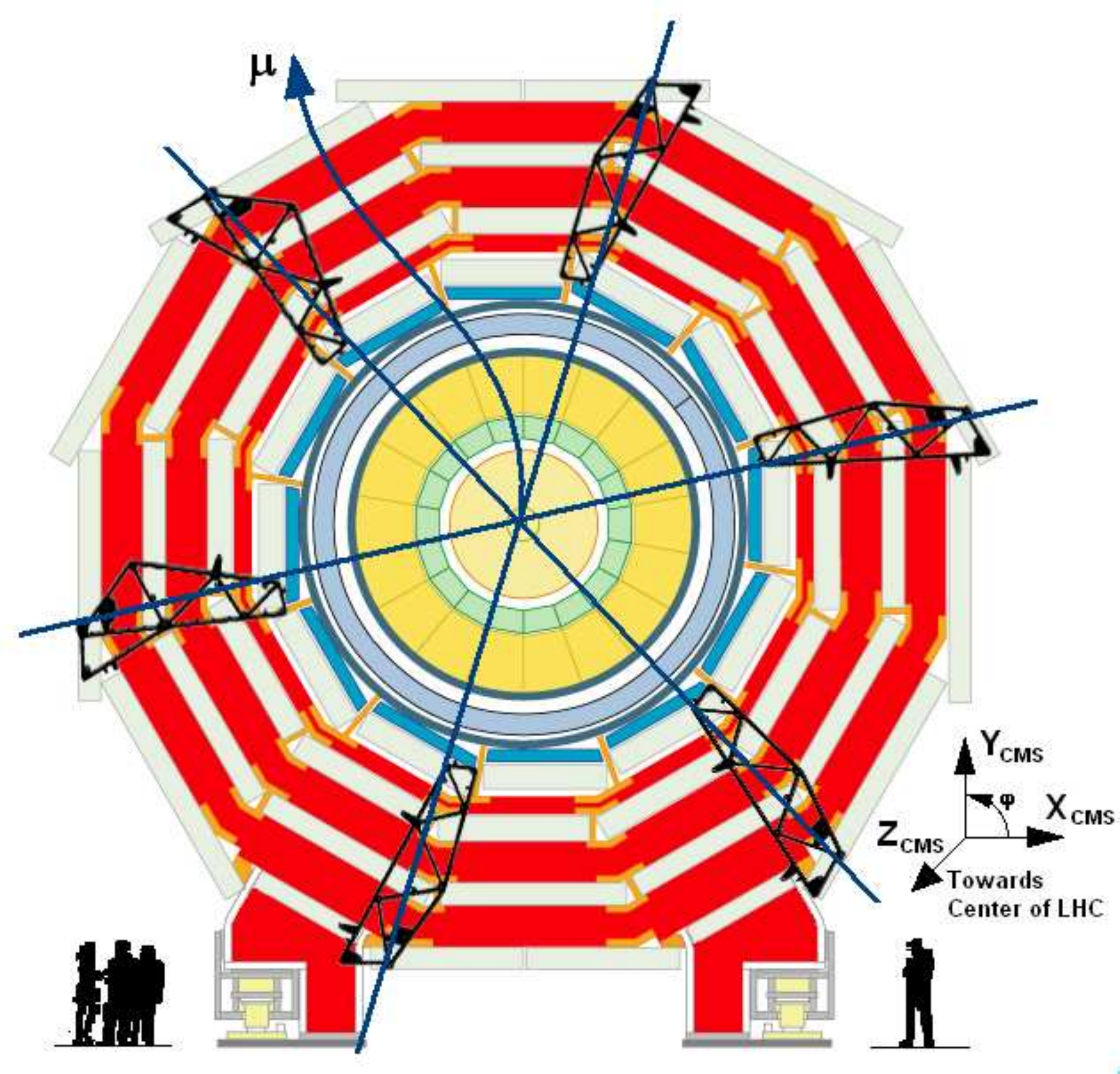}
\caption{Schematic view of the CMS alignment system. Left: longitudinal 
view of CMS showing one of the three  $r_{CMS}$--$z_{CMS}$ alignment planes. The continuous 
and dashed lines show different optical light paths. MABs on YB+1 and YB-1 are not shown
because they are rotated $30^{\circ}$ with respect to this plane. 
Right: transverse view of 
the barrel muon detector. The crossing lines indicate the $r_{CMS}$--$z_{CMS}$ alignment 
planes with $60^{\circ}$ staggering in $\phi$. The $\phi$ sectors are numbered from 1 to 12 
with increasing azimuthal angle, with sector 1 perpendicular to $x_{CMS}$.}
\label{fig:cms_long_trans}
\end{center}
\end{figure}

The basic geometrical layout of the muon alignment system
consists of three $r_{CMS}$--$z_{CMS}$ alignment planes 
with $60^{\circ}$ staggering in $\phi$. This segmentation is based on the 
12-fold geometry of the barrel muon detector (each $\phi$ segment of the DT system 
is called a sector). Within each plane, distributed
networks of optical sensors attached directly to muon chambers or to rigid 
local structures are connected by laser or LED lines. The optical alignment network is 
complemented by different types of analog sensors: electrolytic clinometers, 
optical and mechanical proximity sensors, probes for magnetic field measurement, temperature and 
humidity sensors.
Figure~\ref{fig:cms_long_trans} shows schematic longitudinal and transverse 
views of CMS, with the light paths indicated. 

The muon alignment system comprises three subsystems: endcap, barrel and link.
\\
Each muon endcap station is monitored through three radial straight line monitors running along the full diameter
of the supporting disks. Each straight line monitor consists of laser beams detected by two optical sensors 
in each of the four crossed chambers.
Approximately one sixth of the CSC chambers are directly monitored; 
the rest are aligned with respect to these monitored chambers by detecting tracks 
that pass through their overlapping regions. More details are given in Section 5.
\\
In the barrel, the positions of the 250 DT chambers are monitored by a network of 36 rigid carbon fibre
structures called Modules for the Alignment of the Barrel (MAB), supported by the yoke wheels, and optically 
connected together once the detector is closed. 
Six MABs are mounted on each of the two outermost faces of the external
wheels (see Fig.~\ref{fig:cms_long_trans} right) and inside each of the 4 gaps between the wheels.
The external MABs contain link and endcap elements used to refer the three alignment subsystems to each other.
\\
The Link system simultaneously monitors the twelve external MABs and twenty four 
CSC chambers located in the first endcap 
muon station, in YE$\pm$1 (twelve per side), and relates them to the tracker volume using laser beams emitted 
from two rigid carbon fibre structures, called alignment rings,
attached to the tracker endcaps. Other sources and optics housed in MABs and link disks complete the light path 
layout. The link disks are floating structures suspended inside the innermost endcap iron disks (YE$\pm$1).     

Details on the performance requirements, as well as the design of the subsystems and sensors, can be found 
elsewhere~\cite{MuonTDR,CMS_detector_paper}. Each subsystem performs an independent reconstruction, 
with the link providing a common reference frame. Throughout this paper, geometry reconstruction refers to
the determination of the muon chamber positions and orientations. 

\subsection{Offline geometry reconstruction}
The muon alignment system uses a dedicated reconstruction program
called CMS Object-oriented Code for Optical Alignment (COCOA)~\cite{COCOA1, COCOA2, COCOA3}
to transform the various measurements into a reconstructed DT and CSC aligned 
geometry.
The software reconstructs the position and orientation of the 
optical system objects and chambers, and performs a full propagation of errors to take into account the correlations
between different measurements. 
For the entire muon alignment system, COCOA works with about 
30\,000 parameters: $\approx 3000$ for the link, $\approx 6500$ for the endcap, and
$\approx 20\,000$ for the barrel.
The alignment geometry of the chambers 
and all alignment objects within the system are organized in a hierarchical order using a 
system description which must be provided in addition to the measurements 
themselves.
This description includes the interconnection of elements, e.g.,\ laser-sensor
association, and the system hierarchy, e.g.,\ system elements association to
mechanical structures, together with an approximation of the 
geometry obtained from other measurements (calibrations or photogrammetry). 
Supplying a good estimate of the initial geometry is not necessary, but speeds up the 
convergence, ensures good quality of the fit results, and helps to avoid falling into 
local minima. 


The starting geometry at B = 0\,T uses two types of photogrammetry and optical survey measurements:
\begin{itemize}
\item photogrammetry measurements of DT and CSC chambers, and alignment components inside the disks and wheels. 
These measurements are done with an open detector with respect to the disks and wheels themselves (regardless
of their position and orientation). The measurement precision per point measured is 300\,$\mu$m in all three 
coordinates. 
\item Survey measurements, generally performed with theodolites, of the large structures such as wheels and disks 
after closing the detector  
and reaching the final position. Theodolite measurements have a similar precision in the measured points as photogrammetry, 
but the measurement is done in the global CMS reference and in this transformation the precision is lower, 
so the final coordinates of the large structures have a precision of $\approx 1$\,mm. 
\end{itemize}
%
The geometry reconstruction proceeds independently for 
each alignment subsystem, each having a completely different set of measurements and
geometry description, and thus requiring a different fit strategy implemented in COCOA.
The output of COCOA contains the best geometrical
description of the system compatible with the measurements and with the information from structure calibrations.
Propagated uncertainties for all aligned objects are also provided. 

\subsection{Performance and validation}

Alignment results must be validated before they can be used for track reconstruction. 
For central Drift Tubes, several well established track-based validation techniques are 
described in Ref.~\cite{CRAFT08_track_alignment_paper}. This validation has so far only been performed
for the track-based alignment. For endcap Cathode Strip Chambers, 
the validation procedures are still being developed. 

For the optical-based measurements presented in this paper, the accuracy is 
estimated by comparing the results from geometry reconstruction at 0\,T  with photogrammetry 
measurements. This is the only independent reference used so far to cross-check the results. 
All the comparisons are done independently for each structure, which is considered more stable 
under field-induced deformations.   
In the case of the barrel, the central barrel wheel (YB0) is generally used 
as reference because it is expected to be the most stable structure in CMS.

Care must be taken when comparing photogrammetry measurements, which are taken with an open detector, 
to alignment measurements after detector closing and 
before any magnet cycles which can cause permanent movements and deformations of large structures.
Such measurements are not always directly comparable, since individual
measurements might be biased by residual deformations caused by detector lowering (from the surface to 
the underground cavern), magnetic and 
gravitational forces, internal deformations during closing, or thermal effects.
It is assumed that, in average, the photogrammetry values and the reconstructed values agree.
Under this assumption and in the absence of systematic biases in the reconstruction, 
the distribution of the difference between photogrammetry and reconstructed positions is expected to have a 
mean of zero. The deviation from zero is taken as an estimate of the systematic error 
in the reconstruction.

Until a more precise track-based validation is implemented, the precision of the system is 
given by the standard error propagation of the geometry reconstruction software, 
which can be checked from residual distributions of laser hits with the 
corresponding pull distributions.

\section{Data Acquisition}
The muon alignment data acquisition system is independent of the CMS event flow.
Each alignment subsystem consists of entirely different types of sensors and
electronics. The time required to record a complete set of data for each subsystem is 
$\approx$~18 minutes for the endcap, $\approx$~27 minutes for the link, and $\approx$~2 hours
for the barrel. The data acquisition program is fully integrated 
into the CMS detector control system. The data are recorded in a dedicated online database,  
converted into ROOT~\cite{ROOT} format, and transfered to the Tier-0 computing centre at CERN.
This process was not automated during CRAFT. 
From the Tier-0, the data are copied to an offline farm where geometry reconstruction is performed. 

The knowledge of detector conditions, such as the magnetic field, is of particular importance for the
alignment system. The magnetic field cycles during the entire CRAFT period are shown in 
Ref.~\cite{CRAFTGeneral}. The current data taking strategy 
is to record a full set of alignment measurements at least once per day, under stable conditions. 
Whenever the magnet ramps up or down, information is recorded only from 
analog sensors, such as clinometers and proximity sensors, which have an
almost instantaneous response, in order to track detector movements and deformations.
As the optical sensor cycles are too slow compared to normal ramping times,
full data recording during ramps is not practical. Online monitoring tools are 
being developed in order to provide live information of relative movements of
structures directly monitored by analog sensor measurements and,
with a maximum delay of two hours, information of relative changes 
from optical sensors. 

The full muon alignment system was exercised for the first time during CRAFT. 
The data acquisition and transfer, though not entirely automated, were robust and performed
without any significant operational problems. Data were systematically recorded during
the entire CRAFT period at stable magnetic field values of 0~T and 3.8~T. 

A high operational efficiency was achieved. Above 98\% of optical sensors were operational for the whole 
system. System failures were mainly related to laser misalignment. As an example, the closing 
outside of tolerances of the innermost endcap disk on the $-z_{CMS}$ side 
caused a conflict with some of the link alignment components, 
making the laser system for this part of the detector effectively 
unusable. Most hardware problems were fixed between the subsequent opening 
and closing of the detector.


\section{Muon Barrel Alignment}
The alignment system~\cite{MuonTDR,CMS_detector_paper} measures the positions of the DT 
chambers with respect to each 
other and to the entire muon barrel. Each DT is equipped with LED light sources, visible from  
both $z_{CMS}$ directions, which are observed by 
small video cameras (600 in total) mounted on the MABs. MABs located on each of the three alignment planes
look in opposite directions at the LEDs on the chambers of two mobile wheels. 
The MABs between two wheels observe the LEDs of the chambers of both wheels. 
Direct observations between MABs create also diagonal connections, 
adding further redundancy and 
allowing the system to relate measurements from different wheels and sectors. The $z_{CMS}$ coordinate of the 
MABs is also independently measured with respect to the outer cylinder of the coil. In addition, in the outermost 
wheels of the barrel yoke (YB$-$2 and YB+2) the 3D positions of the MABs are also measured, by the link alignment 
system, in a global reference system with respect to the tracker volume, and thus their relative positions with 
respect to the first endcap station are also known.      

Although raw data for the whole barrel muon system were collected, a very selective data quality control is 
required before the raw data can be used for analysis. Automated data quality procedures were not fully developed   
during CRAFT, so a full geometry reconstruction was not possible.
Only four out of twelve sectors were analyzed and, therefore, only partial results (relative motions) 
are presented here.

\begin{figure}[htbp]
\begin{center}  
\includegraphics[width=0.90\textwidth]{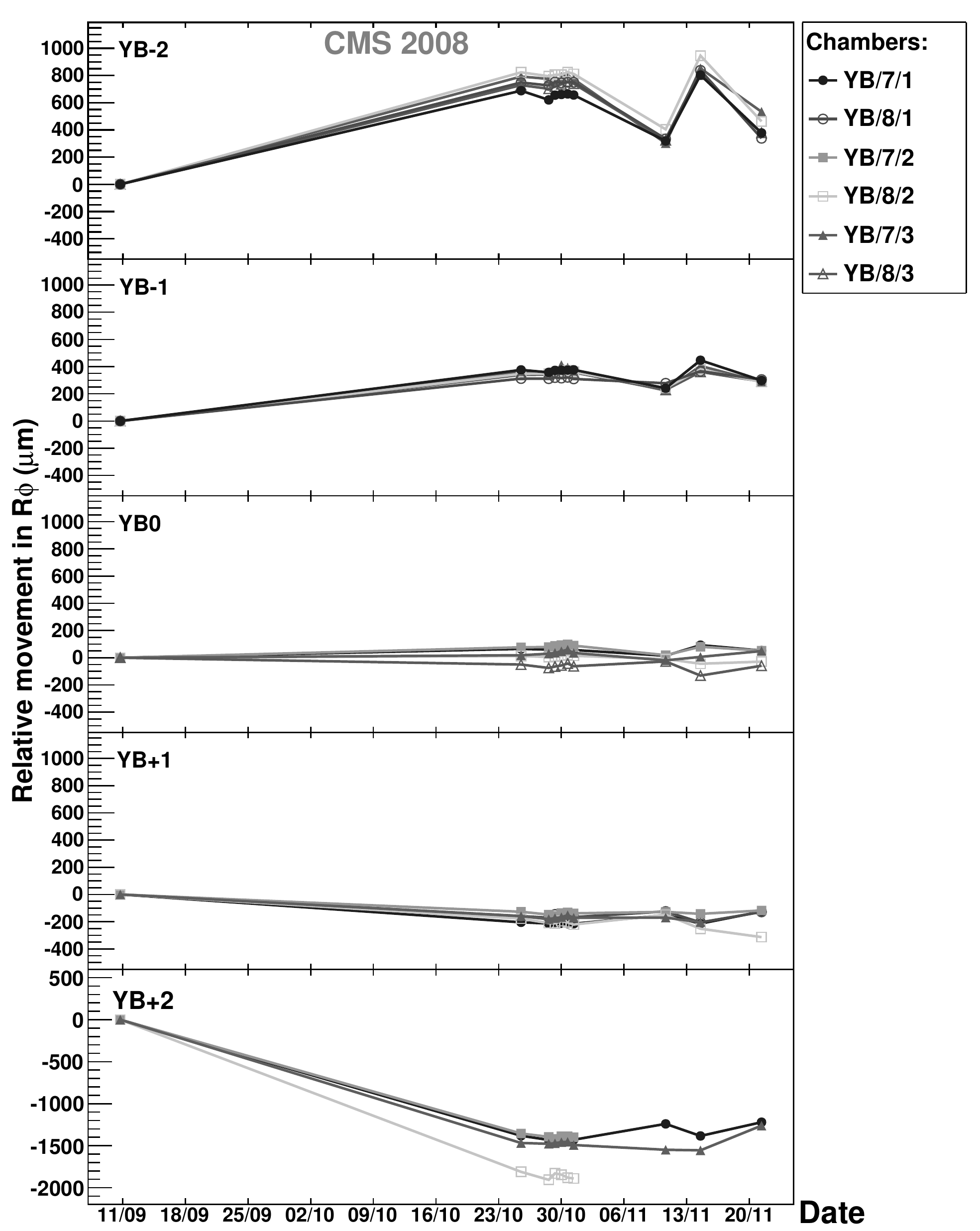}
\caption{Relative $r\phi$ movements of the barrel chambers. Points at 11/09 are the initial 0\,T
measurements before any magnet cycle. The remaining measurements show the difference 
in $r\phi$ with respect to this initial measurement. The cluster of 
points near 30/10 correspond to B~=~3.8\,T. Points near 11/11, 15/11, and 23/11 correspond to B~=~0\,T, 4\,T, and 0\,T, 
respectively. The average precision is $\approx 200\,\mu$m for all points.}
\label{fig:barrelstabfull}
\end{center}
\end{figure}

Using two of the analysed sectors, the stability and reproducibility of the muon barrel geometry is studied by 
comparing data taken at different magnetic field strengths. Relative 
movements in $(r\phi)_{CMS}$ with respect to the first measurement at 0\,T are shown 
in Fig.~\ref{fig:barrelstabfull} for muon stations 1 through 3 in sectors 7 and 8. 
The typical reconstruction precision in position coordinates is estimated to be 200\,$\mu$m. 
During the first rampings of the magnet, large irreversible movements were observed.
Movements in $r\phi$ between the initial 0\,T and 3.8\,T are approximately
0.7, 0.4, 0.1, $-0.2$, and $-1.5$\,mm for wheels YB$-$2, YB$-$1, YB0, YB+1 and YB+2, respectively.
Corresponding movements between the initial 3.8\,T and subsequent 0\,T runs are smaller:
$-0.3$, $-0.1$, $-0.1$, $-0.1$ and 0.3\,mm.
The reproducibility between 0\,T positions after the initial 3.8\,T magnet cycle is within 250\,$\mu$m.
The barrel stability at a constant magnetic field of B = 3.8\,T was also studied over a period of one week.
In general, the detected movements are below 100\,$\mu$m, therefore smaller than the estimated reconstructed 
precision. 

In order to validate the measurements of the system, the DT positions reconstructed at 0\,T  are compared 
to the positions measured directly by photogrammetry~\cite{SurveyBarrelIII}. The comparison was made in a 
coordinate system defined by the DT in sector 7, station 2, considered as perfectly measured by both methods. 
The positions of the other chambers were determined with respect to this one. Only DT chambers in YB0 were used because 
the photogrammetry data for the other wheels include uncertainties of the wheel positions with respect to YB0
and the cumulative error is too large to validate the analysis results. The average error of the 
differences in position (orientation) is 1.5\,mm (1.5\,mrad), which are the convolution of photogrammetry 
and COCOA reconstruction uncertainties. In the absence of systematic biases, DT reconstructed positions
should agree with photogrammetry within these ranges. The observed differences are well within 
the quoted uncertainties. 

During a test of the magnet~\cite{MTCC,CMS_detector_paper} carried out in 2006, the bottom part 
of the barrel alignment system was instrumented. With magnetic field forces acting on the detector, 
a compression of the wheels in $z_{CMS}$ towards the interaction point was observed. This compression was 
measured for most of the barrel during CRAFT. The relative changes of the barrel length at different positions 
were measured by the barrel and link subsystems with respect to two stable references: the outer cylinder of the 
coil and the tracker volume. The results are summarized in Table~\ref{tab:barrelshrink}. 
The compression is not uniform and depends on 
the azimuthal angle, with maximum compression observed at the bottom of the barrel and decreasing towards the top.
For the first two rows in the table, the uncertainty of these measurements is 300\,$\mu$m. The table also shows 
the compression measured independently by the link alignment from a 3D reconstruction of the MABs located on the 
external YB+2 wheel. The uncertainty in these measurements is 280\,$\mu$m, resulting from the convolution of 
the uncertainties in the reconstruction at B = 0\,T and 3.8\,T. 

\begin{table}[htbp]
\caption{Compression of the barrel at B = 3.8\,T compared to B = 0\,T at azimuthal positions of the $z$-bars. 
Missing entries are due to hardware failures. The last line corresponds to the compression measured independently by the link alignment.}
\begin{center}
\scalebox{0.8}{%
\begin{tabular}{|l|c|c|c|c|c|c|} \hline
$\phi_{CMS}$ position $[^{\circ}]$            & 15   & 75 & 135  & 195  & 255  & 315  \\ \hline
$\Delta(z_{CMS})$ at $+z_{CMS}$ side [mm] & $-1.7 \pm 0.3$ &  - & $-1.5 \pm 0.3$  & $-1.7 \pm 0.3$  & $-2.7 \pm 0.3$  & $-2.2 \pm 0.3$  \\ \hline
$\Delta(z_{CMS})$ at $-z_{CMS}$ side [mm] & $1.3 \pm 0.3$  & $1.1 \pm 0.3$ & $1.5 \pm 0.3$ & $1.6 \pm 0.3$ & $1.8 \pm 0.3$ &  -   \\ \hline
Link $\Delta(z_{CMS})$ at $+z_{CMS}$ side [mm] & $-1.8 \pm 0.3$  & $-1.1 \pm 0.3$ & $-1.0 \pm 0.3$  & $-1.8 \pm 0.3$  & $-3.2 \pm 0.3$  & $-2.7 \pm 0.3$  \\ \hline
\end{tabular}}
\end{center}
\label{tab:barrelshrink}
\end{table}

\section{Muon Endcap Alignment}
%
The alignment system monitors the positions of a subset of the 468 Cathode
Strip Chambers in the two muon endcaps. The system uses transverse laser lines
across the face of each yoke disk to measure the deflection in $z_{CMS}$
and longitudinal laser lines to measure the rotation of each yoke disk.

A complex arrangement of different types of position sensors
measures the global $r_{CMS}$, $\phi_{CMS}$, and $z_{CMS}$ coordinates of one-sixth of all CSC chambers.
This allows adequate monitoring of the yoke disk deformations due to strong magnetic forces.
Unmonitored CSC chambers can be coarsely aligned with the average displacements and rotations
observed for monitored chambers in the corresponding rings. All CSC alignments are subsequently refined with
tracks that traverse overlapping chambers. For the ME1/3 chambers, which do not overlap,
azimuthal distances between the chambers are additionally monitored by proximity sensors.

\begin{figure}[htbp]
\begin{center}
\includegraphics[width=0.8\textwidth]{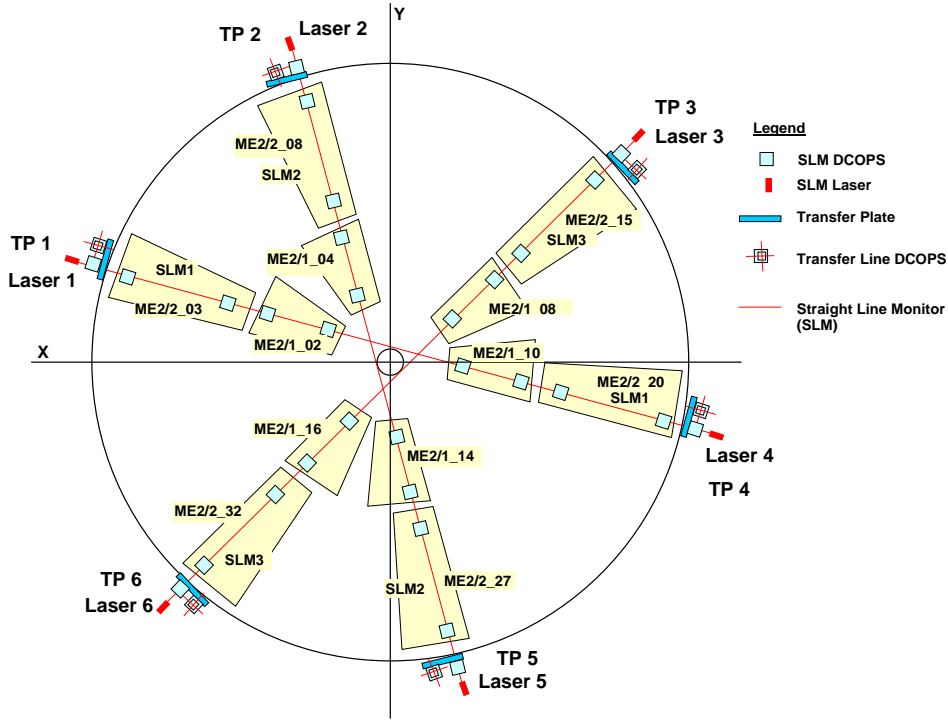}
\caption{Schematic $r\phi$ view of Straight-Line Monitors in the ME2 station.
Locations of axial transfer lines running perpendicular to the plane and across endcaps are
indicated. Optical sensors and other alignment components are also shown.}
\label{fig:Endcap_Align}
\end{center}
\end{figure}

Details on the performance requirements and the design of the system and its sensors can be found
elsewhere~\cite{MuonTDR, CMS_detector_paper,IEEE2006}. Here we briefly recall the main features
of the system used in this analysis. Three straight-line monitors (SLMs) within the $r\phi$ plane of each muon
endcap station, as shown in Fig.~\ref{fig:Endcap_Align}, measure global $z_{CMS}$ and $\phi_{CMS}$ chamber
positions relative to references located at the outer edges of the stations.

In the ME1 stations, the straight-line monitors cannot reach across the entire endcap yoke disk
as they are blocked (by design) by the endcap calorimeters attached to the YE1 yoke disks.
Instead of three full-length straight-line monitors, there are six half-length straight-line monitors
in each ME1 station. Each half straight-line monitor observes two chambers
(one in the ME1/2 ring and one in the outer ME1/3 ring) and connects to the link alignment
laser lines at the inner radius and to the Barrel MABs at the outer radius.
The geometry of the ME4 straight-line monitors is identical to those on the ME2,3 stations despite
the absence of the outer ring of ME4/2 chambers. This identical design is intended
as a preparation for a future planned CMS upgrade that will add outer ME4/2 chambers.

Optical transfer lines run parallel to the CMS $z$-axis along the
outer cylinder envelope of CMS, at six angles separated by $60^{\circ}$ in $\phi_{CMS}$.
These axial lines provide an optical connection between the forward and backward muon
endcaps, across the barrel wheels. Distancemeters, mounted around the periphery of each ME station,
measure the $z_{CMS}$ location of the outer edge of the ME stations relative to the external MAB positions on
the YB2 wheels.

\subsection{Reconstruction of CSC positions}

The analysis of CRAFT data focuses on determining CSC chamber positions in global $z_{CMS}$ and
rotations $\phi_{x_\textrm{local}}$
around their local $x$-axes. This complements track-based alignment, which is mainly suitable for
alignment in the $(r\phi)_{CMS}$ plane~\cite{CRAFT08_track_alignment_paper}.

\subsubsection{Chamber positions in stations ME2,3,4 measured with straight-line monitors}

\begin{figure}[htbp]
\begin{center}
\includegraphics[width=0.5\textwidth,angle=90]{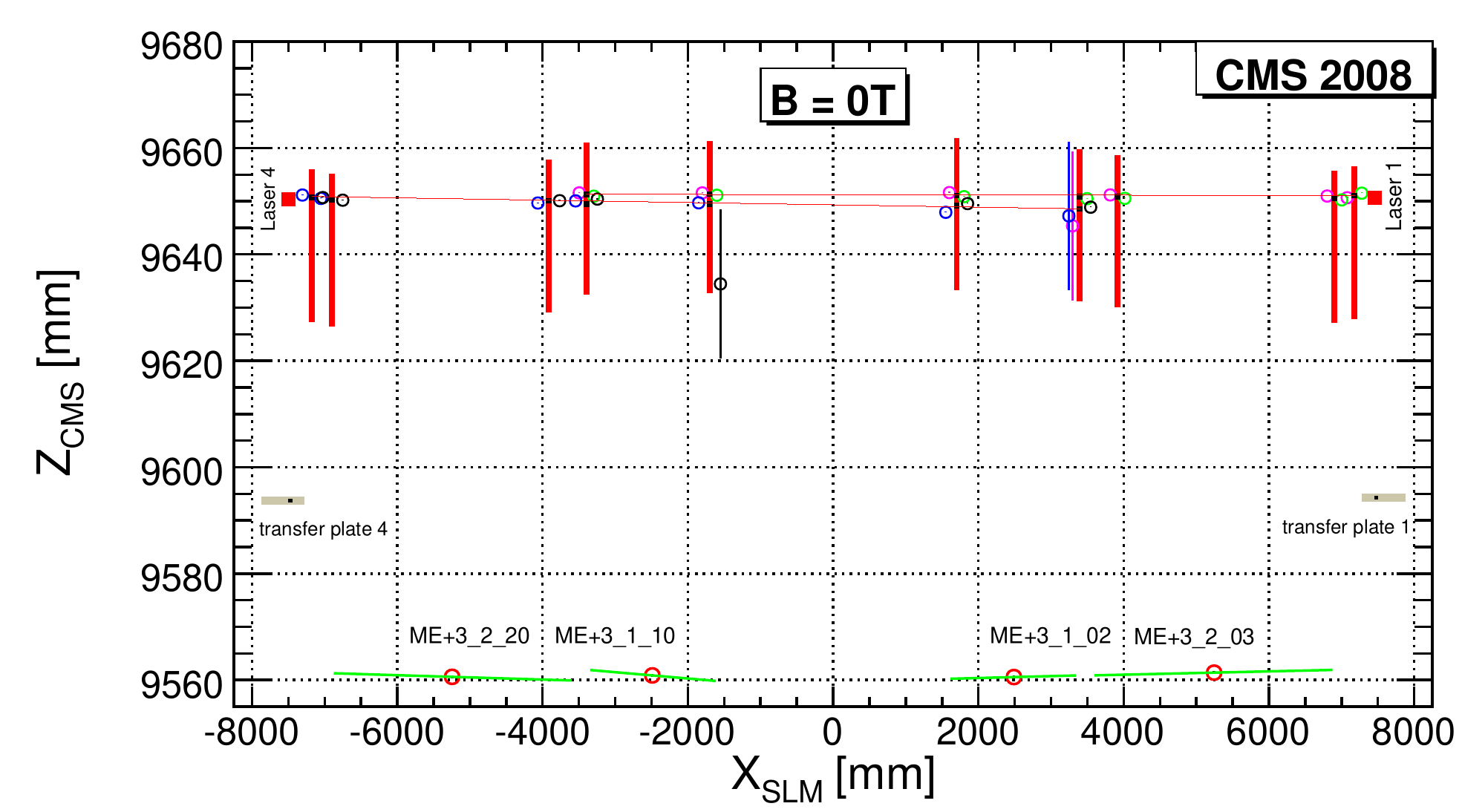}
\includegraphics[width=0.5\textwidth,angle=90]{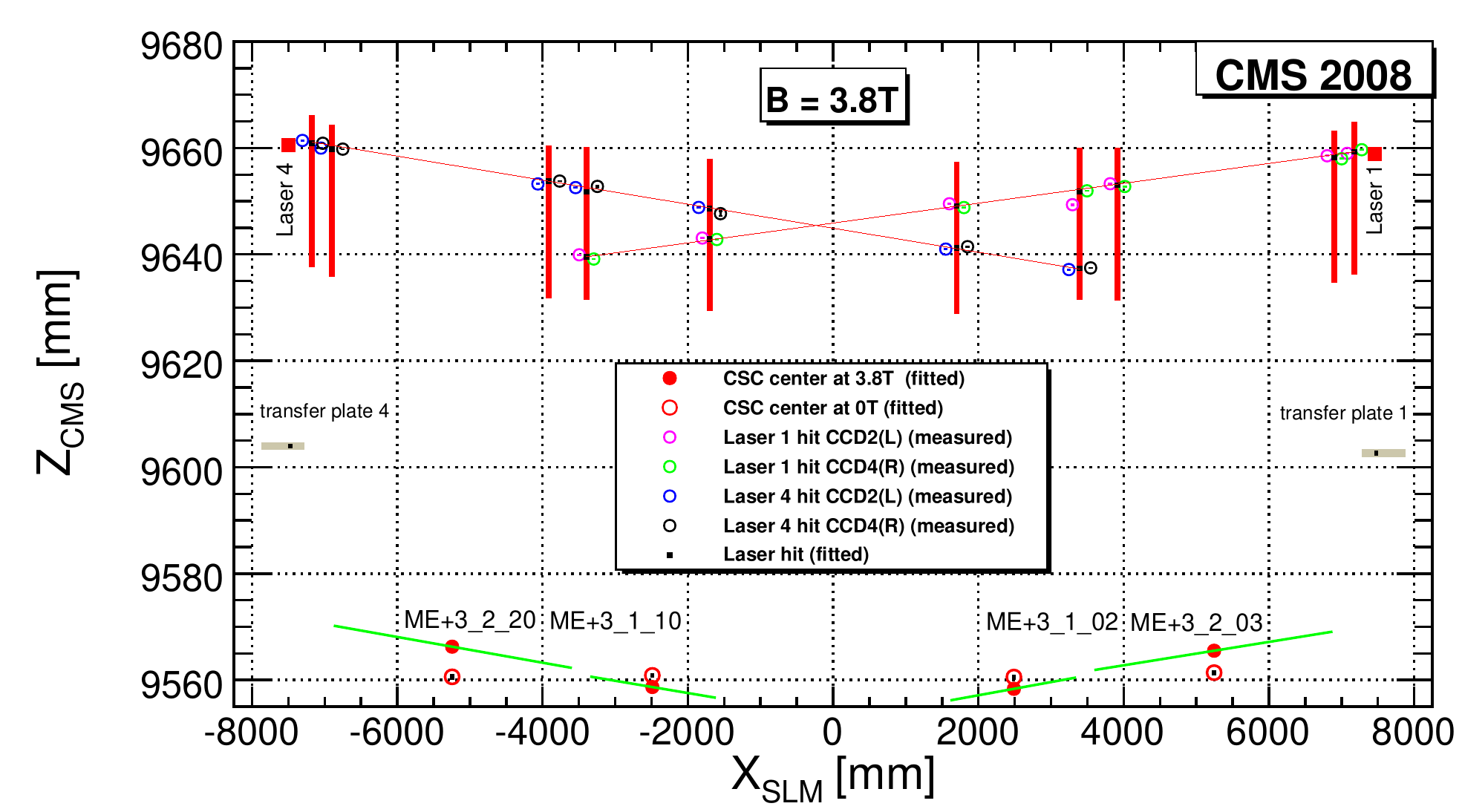}
\caption{Reconstruction results for positions of chambers and sensors in $z_{CMS}$ vs.\ their positions $x_{SLM}$ along a
straight-line monitor at B = 0\,T (top) and 3.8\,T (bottom). The data shown are for the straight-line monitor
connecting transfer plates 1 and 4 of station ME+3. Large vertical bars indicate fitted  positions of
optical sensors. Laser hits on optical sensors are indicated by open circles
with error bars on hit positions  smaller than marker symbols. Long near-horizontal lines represent
fitted laser beams. Reconstructed CSC chamber tilts are indicated by the short near-horizontal lines.}
\label{fig:Endcap_Reconstruction_SLM}
\end{center}
\end{figure}

\begin{figure}[htbp]
\begin{center}
\includegraphics[width=0.32\textwidth,angle=90]{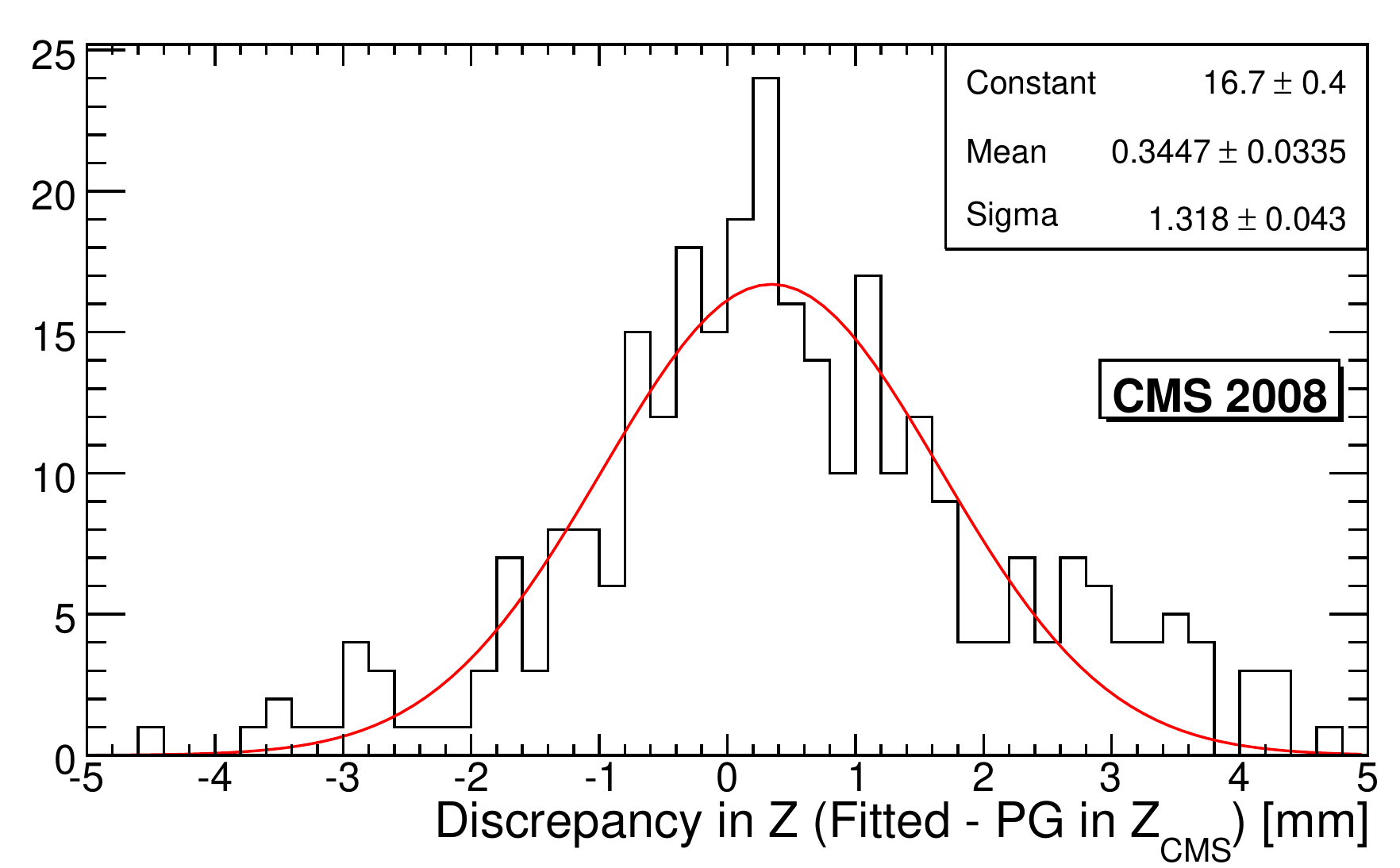}
\includegraphics[width=0.43\textwidth]{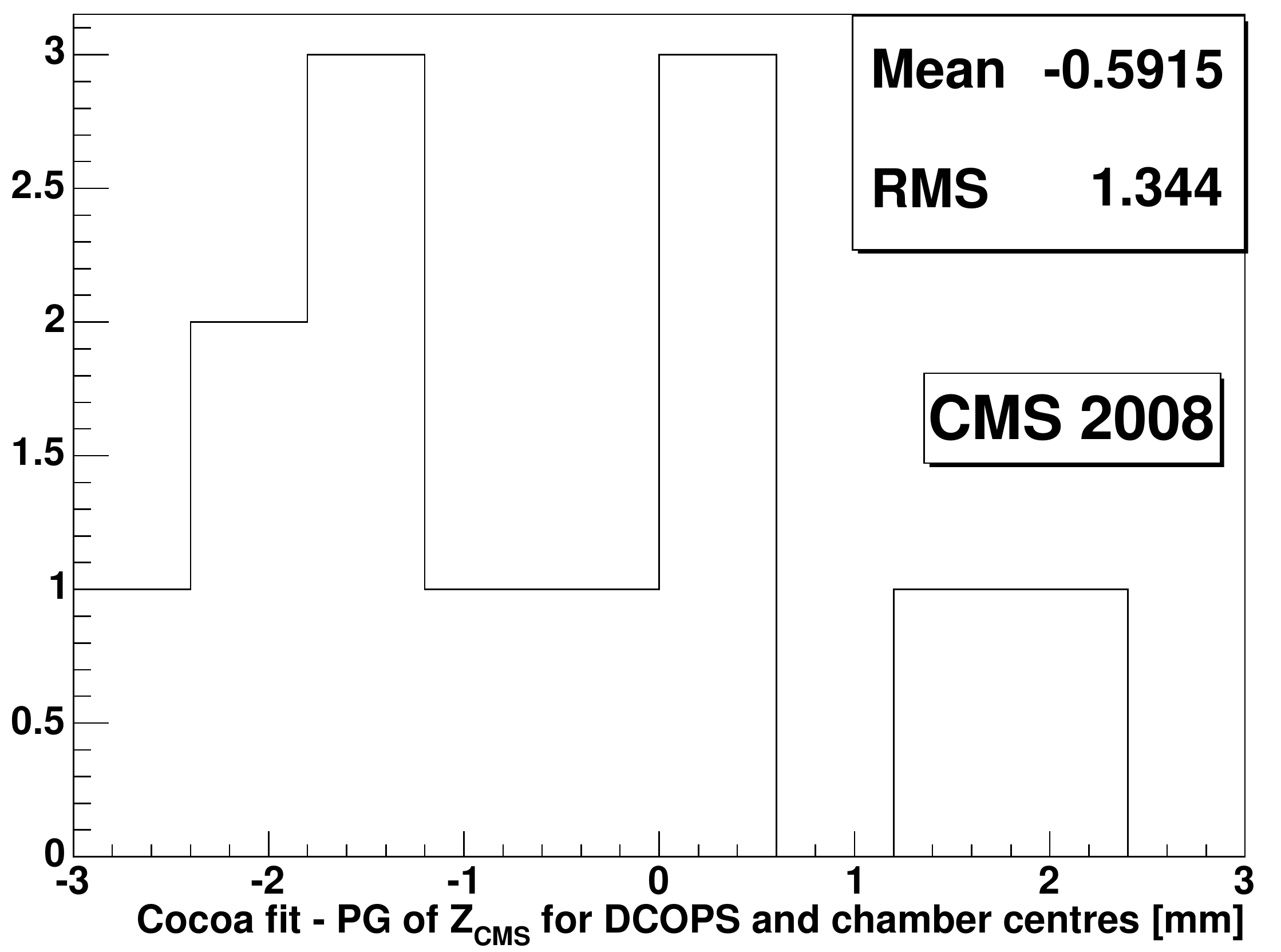}
\caption{Discrepancies $\Delta z_{CMS}$ between $z_{CMS}$-positions reconstructed by COCOA and
measured by photogrammetry (PG) at B = 0\,T. Left: For CSC centres, alignment pins, and optical sensors
in ME2,3,4 stations using full straight-line monitors fitted with a Gaussian.
Right: For CSC centres and optical sensors in ME+1 station using half straight-line monitors.}
\label{fig:Endcap_Reco_vs_PG_DeltaZ}
\end{center}
\end{figure}

The geometry reconstruction starts with performing the basic alignment
procedure described in Section 2, which is common to all subsystems.
Measurements taken at B = 0\,T are used for the reconstruction of
CSC positions. An example of such a reconstruction in $z_{CMS}$ at B = 0\,T is
shown for one straight-line monitor in Fig.~\ref{fig:Endcap_Reconstruction_SLM} (top). The
discrepancies in $z_{CMS}$ between reconstructed values and photogrammetry measurements are
shown in Fig.~\ref{fig:Endcap_Reco_vs_PG_DeltaZ}~(left) for all ME2,3,4 stations.
After the comparison with photogrammetry, the endcap reconstruction proceeds to reconstruct the absolute
$z_{CMS}$ positions of all monitored CSC chambers at B = 3.8\,T, using as
input field-on sensor measurements, the absolute $z_{CMS}$ positions of the endcap yoke disk centers provided by
survey~\cite{SurveyEndcapII} and listed in Table~\ref{tab:tabSurvey}, and the relative $z$-displacements measured
between B = 0\,T and B = 3.8\,T.
\begin{table}[htbp]
\caption{Positions of muon endcap yoke disk centers in global CMS coordinates, measured by survey
with closed and locked detector before the start of the CRAFT test. Nominal global $z$-position
of yoke disk centers are also shown for comparison (nominal (x,y) coordinates are (0,0) for all yoke disks.
The uncertainty for all listed measurements is 0.3\,mm.}
\begin{center}
\scalebox{1.0}{%
\begin{tabular}{|l|c|c|c|c|} \hline
\bf{Yoke disk center } & $x_{CMS}^\textrm{meas}$ [mm] & $y_{CMS}^\textrm{meas}$ [mm] & $z_{CMS}^\textrm{meas}$ [mm]& $z_{CMS}^\textrm{nominal}$ [mm]  \\ \hline \hline
\bf{YE+3}    &  1.6   &  0.6  &  9906.5 &  9900.0 \\ \hline
\bf{YE+2}    &  0.2   &  0.8  &  8828.3 &  8820.0 \\ \hline
\bf{YE+1}    &  1.5   & $-0.3$  &  7568.2 &  7560.0 \\ \hline
\bf{YE--1}    &  2.0   & $-0.4$  & $-7561.7$ & $-7560.0$ \\ \hline
\bf{YE--2}    &  4.1   & $-1.1$  & $-8821.6$ & $-8820.0$ \\ \hline
\bf{YE--3}    & $-1.1$   &  0.4  & $-9903.1$ & $-9900.0$ \\ \hline
\end{tabular}}
\end{center}
\label{tab:tabSurvey}
\end{table}
As an example of field-on reconstruction results,
Fig.~\ref{fig:Endcap_Reconstruction_SLM}~(bottom) shows the reconstruction for the straight-line monitor
between transfer plates 1 and 4 in ME+3 at B = 3.8\,T. Comparing this result with the reconstruction for
the same straight-line monitor at B = 0\,T (Fig.~\ref{fig:Endcap_Reconstruction_SLM}~top) demonstrates that the
endcap yoke disk bends due to magnetic forces. Analogous reconstruction plots for data taken at
B = 3.8\,T with the other straight-line monitors show similar deformations of all yoke disks towards the interaction point
by about 10 to 12\,mm for the chambers on the inner rings and by about 5\,mm away from the interaction point
for the chambers in the outer ring.

\begin{figure}[htbp]
\begin{center}
\includegraphics[width=0.8\textwidth]{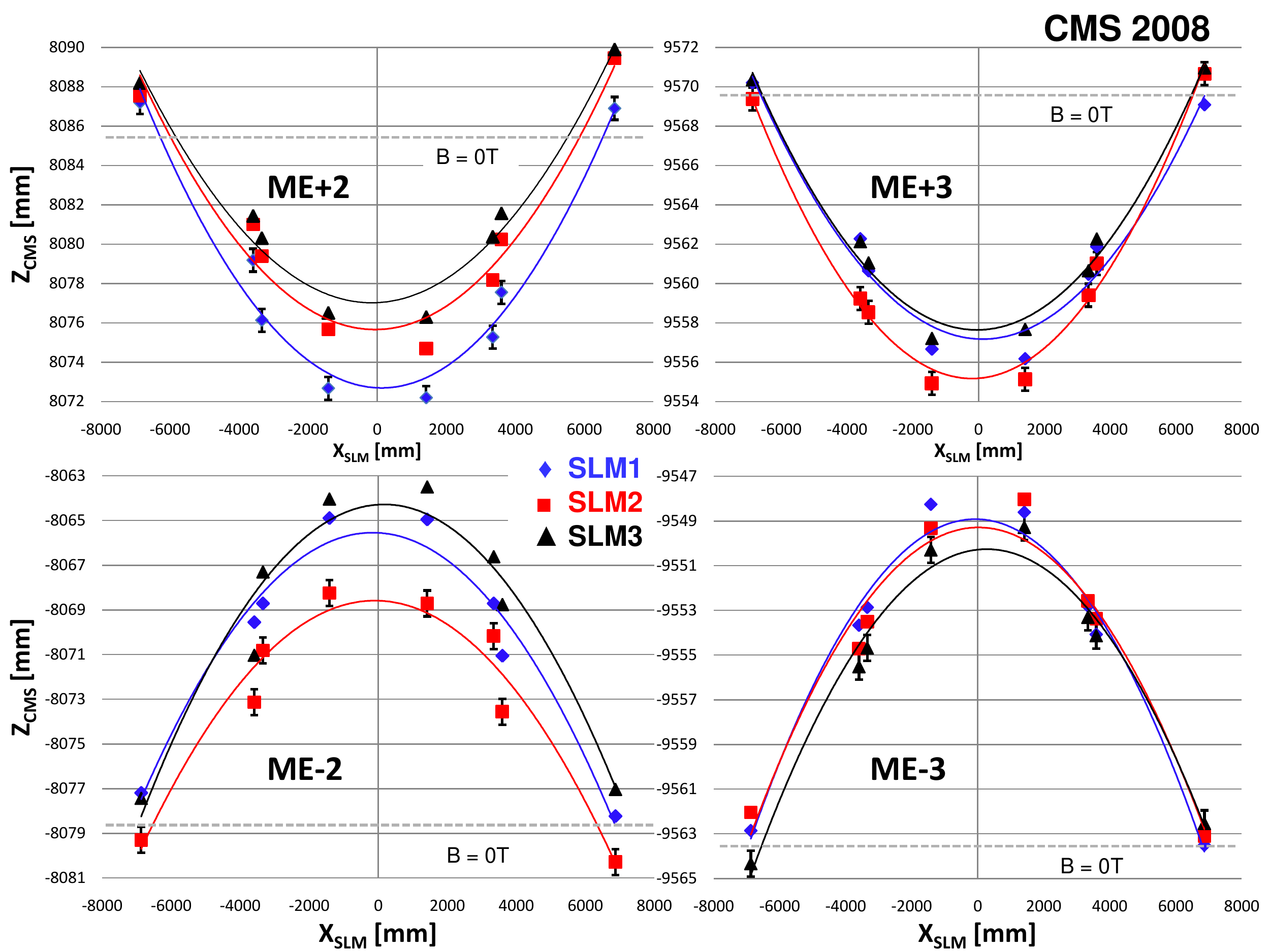}
\caption{Yoke disk deformations at B = 3.8\,T in ME2 and ME3 stations measured with straight-line monitors.
Points shown correspond to positions of CSC alignment pins and are fitted with second order polynomials.
Error bars correspond to $\sigma_{tot}(z_{CMS}) = 470\,\mu$m and apply to all measurement points. For clarity,
error bars are shown only for the lowest curve in each station.
shed lines indicate the average alignment pin positions at B = 0\,T
r the three SLMs in each station as measured by photogrammetry and survey.}
RESPONSE TO JINST REFEREE
\label{fig:ME_disk_bending}
\end{center}
\end{figure}

We summarize our reconstruction results for the ME2,3 stations in Fig.~\ref{fig:ME_disk_bending}.
The yoke disk deformations shown in this figure are calculated from the differences of
reconstructed positions of the CSC alignment pins at B = 0 and 3.8\,T. Quadratic fits describe
the yoke disk deformations reasonably well. These results are consistent with ME+2 reconstruction results
using first magnet test data taken in 2006 at 4\,T~\cite{IEEE2007}. The differences observed by
different straight-line monitors for the same station indicate a slightly asymmetric
deformation of the yoke disks.
Some asymmetry is expected because the endcap yoke disks are fixated at the bottom to massive carriage structures
that are used to move the yoke disks as needed.

The precision of the reconstruction is estimated from COCOA's error propagation to be
280\,$\mu$m and 320\,$\mu$m in $z_{CMS}$ for the inner and outer rings of chambers,
respectively, and 200\,$\mu$rad in $\phi_{x_\textrm{local}}$. The angular uncertainty
has a contribution from the error in $z_{CMS}$ and the chamber length.
Figure~\ref{fig:Endcap_Reco_vs_PG_DeltaZ}~(left) shows the difference $\Delta z_{CMS}$
between $z_{CMS}$-positions reconstructed by COCOA and
measured by photogrammetry at B = 0\,T for CSC centers, alignment pins, and optical sensors
in ME2,3,4 stations using full straight-line monitors. We take the deviation from zero
of the mean of this distribution as a measure of the systematic error of the
COCOA reconstruction, i.e.\ $\sigma_\textrm{syst}(z_{CMS}) = 340\,\mu$m.
The photogrammetry measurements used in this comparison spread over a significant time period
of two years.  
Some measurements were done on the surface and some in the underground CMS cavern.
The deviations are therefore an upper limit. Reconstruction at B = 3.8\,T
cannot be checked against independent survey or photogrammetry because those cannot be
performed for a closed detector when the magnetic field is turned on. Consequently, we explicitly assume that
the COCOA reconstruction of CSC positions at B = 3.8\,T has very similar errors as the B = 0\,T
reconstruction, because the same reconstruction method is applied in both cases.

\subsubsection{Chamber positions in ME1}

The reconstruction of CSC positions in ME1 using half straight-line monitor data is more complex than the
reconstruction of full-length straight-line monitors in ME2,3,4, because additional information
from the link alignment is required. We use the $z_{CMS}$ of MABs reconstructed by the link system,
together with measurements from distancemeters, to locate the ME1 yoke disk in $z_{CMS}$.

For the ME1/1 and ME1/2 chambers, $x_{CMS}$, $y_{CMS}$, and $z_{CMS}$ coordinates as well as the chamber
rotations $\phi_{CMS}$ and $\phi_{x_\textrm{local}}$, around their local $x$-axes, are obtained from the
reconstruction of the link sensors data. The reconstruction procedure is described in Ref.~\cite{MTCC}.
Using global $x,y,z$
coordinate information from link sensors inside the half straight-line monitor, together with
global $x,y$ information from transfer lines, the ME1/3 chambers can be located accurately and the
measurements of the ME1/2 chamber positions provided by the link system can be refined.
For the CRAFT exercise, adequate transfer line information was not available, so
the $x_{CMS},\ y_{CMS}$ coordinates were not reconstructed at B = 3.8\,T, and the alignment
focused on measuring the $z_{CMS}$ coordinates.

Note that six ME1/2 chambers are reconstructed independently by both the link and endcap systems.
The relative field-on to field-off movement for these chambers is in very good agreement,
with an average difference between the two systems below 30\,$\mu$m in $x_{CMS}$ and
$y_{CMS}$, and below 190\,$\mu$m in $z_{CMS}$. Absolute reconstructed positions agree
within 470\,$\mu$m.

The precision of the ME1 reconstruction using the half straight-line monitors is estimated
from COCOA's error propagation. For ME1/2 and ME1/3 chambers the error in $x_{CMS}$ and $y_{CMS}$ is
260\,$\mu$m.  The errors in $z_{CMS}$ are 275\,$\mu$m and $345\,\mu$m for the ME1/2 and ME1/3 chambers,
respectively.
We compare the results of the B = 0\,T fit to photogrammetry measurements for ME+1,
Fig.~\ref{fig:Endcap_Reco_vs_PG_DeltaZ} (right), and take the deviation from zero of
the mean of this distribution as a measure of the systematic error of the COCOA
reconstruction, i.e.\ $\sigma_{syst}^{\mbox{\tiny ME+1}}(z_{CMS}) = 590\,\mu$m.
Following the same procedure for the reconstruction of the link system data, the estimated errors on the
determination of the positions of the ME1/1 and ME1/2 chambers in global $x,y,z$ coordinates
are 140 and 220\,$\mu$m for B = 0\,T and B = 3.8\,T fits respectively. COCOA propagation of
angular uncertainties estimates an error of 30\,$\mu$rad for angular
measurements. By comparing the results of the B = 0\,T fit to photogrammetry measurements
for ME+1 we measure a systematic bias at the level of 35\,$\mu$m for positions and 35\,$\mu$rad for angles.


\subsection{Endcap alignment constants}

\begin{figure}[htbp]
\begin{center}
\includegraphics[width=0.8\textwidth]{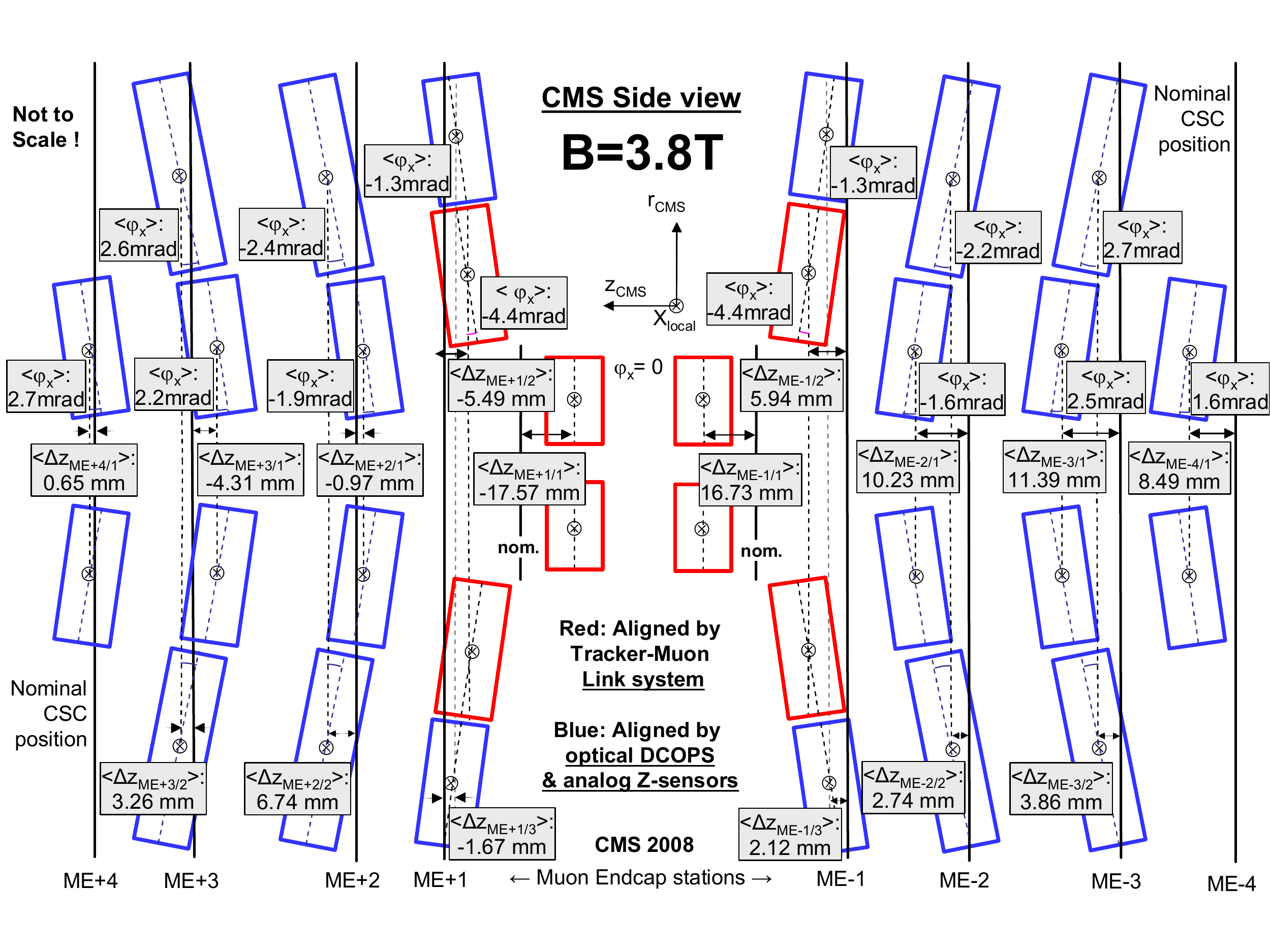}
\caption{Summary sketch of average deformations and displacements from nominal positions for muon endcap stations
at B = 3.8\,T during CRAFT exercise, as observed with the muon alignment system. The shown
displacements $\Delta z_{CMS}$ and rotations $\phi_{x_\textrm{local}}$ are averages over the six monitored CSC
chambers in each ring.}
\label{fig:ME_alignment_overview_sideview_sketch}
\end{center}
\end{figure}

For those CSC chambers that are not directly monitored by straight-line monitors, the average of
the $z_{CMS}$-positions and $\phi_{x_\textrm{local}}$-tilts  obtained from the
monitored chambers are used as alignment corrections in the corresponding rings. This is reasonable
since we find approximate azimuthal symmetry
in the yoke disk deformation (Fig.~\ref{fig:ME_disk_bending}). The average alignment corrections with
respect to the nominal geometry are listed for each CSC ring in Table~\ref{tab:tabConstant} and visualized as a
sketch in Fig.~\ref{fig:ME_alignment_overview_sideview_sketch}. Due to different
initial positions of the forward and backward endcap yoke disks relative to nominal
(see Table~\ref{tab:tabSurvey}) the CSC alignment corrections to $z_{CMS}$ with respect
to {\em nominal} CSC positions are not symmetric between the two endcaps, even though the
bending itself is similar for forward and backward muon stations.
The CSC alignment constants described here together with complementary constants for the
other coordinates ($r\phi$ positions), obtained from track-based muon alignment~\cite{CRAFT08_track_alignment_paper},
are used for the reconstruction of cosmic ray muon tracks in the CRAFT exercise~\cite{CRAFTmuonrecopaper}.
\begin{table}[htbp]
\caption{Average alignment corrections $\Delta z_{CMS} = z_{CMS}^{reco} - z_{CMS}^{nominal}$ to
CSC chamber positions and orientations for each ring with respect to nominal.
The typical precisions are described in the text. Dashes in the table indicate degrees of freedom
not measured by the system.}
\begin{center}
\scalebox{0.9}{%
\begin{tabular}{|l|c|c|c|c|c|c|c|c|} \hline
\bf{Ring  }                                 & ME+1/1 & ME+1/2 & ME+1/3 & ME+2/1 & ME+2/2 & ME+3/1 & ME+3/2 & ME+4/1  \\ \hline \hline
\boldmath{$\Delta z_{CMS}$ [mm]}            & $-17.57$ & $-5.49$ &  $-1.67$  & $-0.97$  &  6.74  &  $-4.31$ &  3.26  &   0.65  \\ \hline
\boldmath{$\Delta\phi_{x_\textrm{local}}$ [mrad]}    &  -- & $-4.4$ & $-1.3$ & $-1.9$   & $-2.4$   &   2.2  &  2.6   &   2.7   \\ \hline \hline
\bf{Ring  }                                  & ME--1/1 & ME--1/2 & ME--1/3 & ME--2/1 & ME--2/2 & ME--3/1  & ME--3/2 & ME--4/1 \\ \hline \hline
\boldmath{$\Delta z_{CMS}$ [mm]}             & 16.73 & 5.94 & 2.12 & 10.23   &  2.74   &   11.39  &  3.86   &   8.49 \\ \hline
\boldmath{$\Delta\phi_{x_\textrm{local}}$ [mrad]}   & -- & $-4.4$ & $-1.3$ & $-1.6$    & $-2.2$    &   2.5    &  2.7    &   1.6  \\ \hline
\end{tabular}
}
\end{center}
\label{tab:tabConstant}
\end{table}

\section{Summary}
The muon alignment system successfully recorded data during the CRAFT 2008 exercise. 
The results obtained show several achievements as well as some shortcomings which are
being or have already been addressed. All three subsystems are able to track
detector movements and deformations under magnetic forces, and monitor the 
stability of the detector during operation. Results from the three subsystems are in 
agreement with each other and with photogrammetry measurements, where these apply.

For the DT chambers, only four (out of twelve) sectors in only one (out of five) wheels
were aligned relative to a fixed DT chamber. Although a complete alignment geometry 
was not achieved, initial partial reconstruction of active planes and studies of 
stability show promising results. Initial reconstructions at B = 0\,T agree with 
photogrammetry within uncertainties. Stability under nominal field 
and reproducibility of reconstructed positions at 0\,T after full magnetic field 
cycles are observed at a level smaller or comparable to the estimated
reconstruction precision of 200\,$\mu$m.
  
For the CSC chambers, all monitored chambers were aligned in $z_{CMS}$ and $\phi_{x_\textrm{local}}$,
with the exception of the ME1/1 chambers, which were aligned only in $z_{CMS}$.  
Alignment of the CSC chambers for the remaining degrees of freedom, in particular for $x_{CMS}$ 
and $y_{CMS}$, which are crucial for momentum reconstruction, is in progress.
An aligned detector geometry at B = 3.8\,T is provided in the form of alignment 
constants which are used for muon track reconstruction. The precision for ME2, ME3, and ME4 chambers
ranges between 280 and 320\,$\mu$m in $z_{CMS}$, and is approximately 200\,$\mu$rad in $\phi_{x_\textrm{local}}$. 
For the ME1 chambers, the precision in $z_{CMS}$ ranges between 220 and 340\,$\mu$m from inner to outer rings. 
The systematic error associated with the reconstruction is estimated to be below 500\,$\mu$m from comparison with 
photogrammetry. 

\begin{table}[htbp]
\caption{Typical precisions obtained for DT and CSC chamber alignment. Dashes in the table indicate degrees of freedom 
not yet measured by the system. Of the reconstructed degrees of freedom, the most relevant for momentum measurement is
$r\phi_{CMS}$, the remaining affecting the momentum reconstruction as a higher-order correction.}
\begin{center}
\begin{tabular}{|l|c|c|c|} \hline
\bf{Chamber  }       & $r\phi_{CMS}$ [$\mu$m] & $z_{CMS}$ [$\mu$m] & $\phi_{x_\textrm{local}}$ [$\mu$rad]  \\ \hline \hline
\ DT            & 200 & --       &  --    \\ 
\ CSC ME1       & --  & 220--340 &  --    \\ 
\ CSC ME2,3,4   & --  & 280--320 &  200   \\ 
\hline
\end{tabular}
\end{center}
\label{tab:tabPrecisions}
\end{table}  

Alignment precisions are summarized in Table~\ref{tab:tabPrecisions}.
The muon alignment system was partly commissioned during CRAFT. 
Even if a complete muon alignment reconstruction was not ready, it has
proved its capability to provide DT and CSC chamber alignment with a precision close 
to that required by CMS. 
Most of the system failures have subsequently been fixed, and initial results 
for a complete alignment of DT and CSC chambers is already available in 2009. 
A very precise and accurate muon alignment is expected in the near future
by combining results from the muon alignment system with those from the
alignment based on tracks.

\section*{Acknowledgments}
We thank the technical and administrative staff at CERN and other CMS Institutes, and acknowledge support from: 
FMSR (Austria); FNRS and FWO (Belgium); CNPq, CAPES, FAPERJ, and FAPESP (Brazil); MES (Bulgaria); 
CERN; CAS, MoST, and NSFC (China); COLCIENCIAS (Colombia); MSES (Croatia); RPF (Cyprus); 
Academy of Sciences and NICPB (Estonia); Academy of Finland, ME, and HIP (Finland); 
CEA and CNRS/IN2P3 (France); BMBF, DFG, and HGF (Germany); GSRT (Greece); OTKA and NKTH (Hungary); 
DAE and DST (India); IPM (Iran); SFI (Ireland); INFN (Italy); NRF (Korea); LAS (Lithuania); 
CINVESTAV, CONACYT, SEP, and UASLP-FAI (Mexico); PAEC (Pakistan); SCSR (Poland); FCT (Portugal); 
JINR (Armenia, Belarus, Georgia, Ukraine, Uzbekistan); MST and MAE (Russia); MSTDS (Serbia); 
MICINN and CPAN (Spain); Swiss Funding Agencies (Switzerland); NSC (Taipei); TUBITAK and TAEK (Turkey); 
STFC (United Kingdom); DOE and NSF (USA). 
Individuals have received support from the Marie-Curie IEF program (European Union); 
the Leventis Foundation; the A. P. Sloan Foundation; and the Alexander von Humboldt Foundation.

\bibliography{auto_generated}   
\clearpage
\cleardoublepage\appendix\section{The CMS Collaboration \label{app:collab}}\begin{sloppypar}\hyphenpenalty=500\textbf{Yerevan Physics Institute,  Yerevan,  Armenia}\\*[0pt]
S.~Chatrchyan, V.~Khachatryan, A.M.~Sirunyan
\vskip\cmsinstskip
\textbf{Institut f\"{u}r Hochenergiephysik der OeAW,  Wien,  Austria}\\*[0pt]
W.~Adam, B.~Arnold, H.~Bergauer, T.~Bergauer, M.~Dragicevic, M.~Eichberger, J.~Er\"{o}
% ...
\vskip\cmsinstskip
% ...

\end{sloppypar}
\end{document}